\documentclass[11pt,a4paper]{article}
\usepackage{jheppub}
\usepackage{amsmath}
\usepackage{amssymb}
\usepackage{graphicx}
\usepackage{babel}
\usepackage[caption=false]{subfig}
\usepackage{mathtools}
\usepackage{slashed}
\usepackage{mathrsfs} 
\usepackage{xcolor}

\newcommand{\ii}{\text{i}}
\newcommand{\ip}{{\bf 7}}
\DeclareMathOperator{\Tr}{Tr}

\definecolor{kblue}{rgb}{0,0.48,0.73}
\definecolor{kred}{rgb}{0.73,0.25,0}
\definecolor{kgreen}{rgb}{0.48,0.73,0}
\definecolor{dgreen}{rgb}{0.2,0.51,0.19}
\definecolor{dred}{rgb}{0.7,0.15,0.09}

\usepackage{ulem}

\begin{document}
\allowdisplaybreaks[4]

\begin{titlepage}

\begin{center}

{\LARGE \bfseries
Fission and molecular binding of\\
non-Abelian \(A_4\) vortices 
in $SO(3)$ gauge theory
}

\vspace{1cm}

\renewcommand{\thefootnote}{\fnsymbol{footnote}}
{%
\hypersetup{linkcolor=black}
Yoshihiko Abe$^{1,2,3}$\footnote[1]{yabe3@keio.jp},
\ 
Tetsutaro Higaki$^{4,5}$\footnote[2]{thigaki@rk.phys.keio.ac.jp},
\ 
Kazuya Murakami$^{4}$\footnote[3]{kazuyamurakami@keio.jp},
\ 
Muneto Nitta$^{5,6,7}$\footnote[4]{mune.nitta@gmail.com},
\ and 
Ryo Yokokura$^{5,6}$\footnote[5]{ryokokur@keio.jp}
}%
\vspace{8mm}

{\itshape%
$^1${Graduate School of Science and Technology, Keio University, Yokohama , Japan}\\
$^2${Keio University Sustainable Quantum Artificial Intelligence Center (KSQAIC), Keio University, Tokyo 108-8345, Japan}\\
$^3${Quantum Computing Center, Keio University, 3-14-1 Hiyoshi, Kohoku-ku, Yokohama, Kanagawa, 223-8522, Japan}\\
$^4${Department of Physics, Keio University, Yokohama 223-8533, Japan}\\
$^5${Research and Education Center for Natural Sciences, Keio University,
4-1-1 Hiyoshi, Yokohama, Kanagawa 223-8521, Japan}\\
$^6${Department of Physics,  Keio University,
4-1-1 Hiyoshi, Yokohama, Kanagawa 223-8521, Japan}\\
$^7${International Institute for Sustainability with Knotted Chiral Meta Matter(WPI-SKCM$^2$), Hiroshima University, 1-3-1 Kagamiyama, Higashi-Hiroshima, Hiroshima 739-8531, Japan}
}%

\vspace{8mm}

\end{center}

\begin{abstract}
~
We investigate the nonlinear evolution of unstable vortices in an \(SO(3)\) gauge--Higgs theory in which the gauge symmetry is spontaneously broken to the tetrahedral subgroup \(A_4\).
Previous work constructed two families of vortices associated with elements of order two and three in \(A_4\), referred to as order-two and order-three vortices, respectively, and found that order-two vortices can be unstable.
Building on this analysis, we numerically study the evolution of vortex configurations perturbed along their unstable modes using gradient flow.
First, when all gauge- and Higgs-boson mass scales are equal, we find that an order-two vortex splits into two order-three vortices.
Second, at a special choice of mass ratios, a doubly wound order-three vortex evolves toward a bound configuration of two order-three vortices at finite separation, whose existence was suggested by the previous stability analysis.
We further confirm numerically that the resulting vortex molecule is linearly stable in the full \(SO(3)\) theory.
\end{abstract}

\end{titlepage}

\renewcommand{\thefootnote}{\arabic{footnote}}
\setcounter{footnote}{0}
\setcounter{page}{1}

\tableofcontents

\section{Introduction}
\label{sec:introduction}

In the early Universe, spontaneous symmetry breaking can lead to the formation of topological defects, with the type of defect determined by the topology of the vacuum manifold \cite{Kibble:1976sj,Mermin:1979zz,Manton:2004tk}.
Among them, cosmic strings are line-like defects that can arise in a wide class of particle-physics models \cite{Vilenkin:1982sh,Hindmarsh:1994re,Vilenkin:2000jqa}.
Cosmic strings were historically considered as possible sources of cosmological density perturbations \cite{Vilenkin:1981iu,Kaiser:1984iv}.
Observations of the cosmic microwave background have since constrained both their contribution to the observed anisotropies and the allowed string tension \cite{Kaiser:1984iv,Bevis:2006mj,Planck:2013mgr,Planck:2015fie}.
Cosmic-string networks can also generate a stochastic gravitational-wave background, providing a possible probe of particle physics and the thermal history of the early Universe \cite{Vachaspati:1984gt,Auclair:2019wcv}.
The observable signatures of cosmic strings depend not only on the properties of an isolated string, but also on how the string network evolves.

The canonical field-theoretic model of a local cosmic string is the Abrikosov-Nielsen--Olesen vortex of the Abelian-Higgs model \cite{Abrikosov:1956sx,Nielsen:1973cs,deVega:1976xbp}.
The interaction between two vortices is attractive or repulsive 
for type-I or II superconductivity, 
in which the mass of the gauge field 
is heavier or lighter than that of  
the Higgs field,
respectively 
\cite{Jacobs:1978ch,Speight:1996px}.
Their boundary allows
critically coupled
Bogomol'nyi-Prasad-Sommerfield (BPS) vortices~\cite{Bogomolny:1975de,Jaffe:1980mj}.

The cosmological evolution of networks of such strings has been studied extensively \cite{Vachaspati:1984dz,Albrecht:1984xv,Hindmarsh:1994re,Martins:1996jp,Vincent:1997cx,Moore:2001px,Hindmarsh:2017qff}.
When strings intersect, they can intercommute and reconnect, while the network also loses energy through loop formation and radiation \cite{Shellard:1987bv,Albrecht:1989mk,Allen:1990tv,Vincent:1997cx,Hindmarsh:2021mnl}.
As a result, the network approaches a scaling regime in which the characteristic string separation grows in proportion to the horizon scale and the number of long strings per horizon volume remains approximately constant \cite{Albrecht:1989mk,Bennett:1989yp,Allen:1990tv,Martins:1996jp,Hindmarsh:2017qff}. 
Vortices can be extended to those in 
a non-Abelian $U(N)$ Higgs model
\cite{Hanany:2003hp,Auzzi:2003fs,Eto:2004rz,Eto:2005yh,Eto:2006cx,Eto:2006uw,Tong:2005un,Eto:2006pg,Shifman:2007ce,Shifman:2009zz,Tong:2008qd}. 
The reconnection probability is still unity for cosmic strings in the non-Abelian Higgs model \cite{Hashimoto:2005hi,Eto:2006db} and thus the same scaling regime can be expected.

This picture can change when the strings carry non-Abelian topological charges \cite{Bais:1980vd,Alford:1990mk,Alford:1990ur,Alford:1992yx,Lo:1993hp,Brekke:1992he,Brekke:1997jj,McGraw:1997nx,Chen:2026jan}.
For example, when two strings with noncommuting charges collide, they cannot always reconnect in the same way as Abelian strings; instead, topological constraints can lead to the formation of connecting bridges or more complicated string webs \cite{Poenaru1977,Mermin:1979zz,Bais:1980vd,Alford:1992yx,McGraw:1997nx,Kobayashi:2008pk,Brekke:1992he}.
Such effects can modify both the structure and the scaling properties of the network \cite{McGraw:1997nx,Chen:2026jan}.
In condensed matter physics, 
non-Abelian topological defects have also been studied in
biaxial liquid crystals~\cite{Poenaru1977,Mermin:1979zz},
chiral nematic systems with effective biaxial symmetry \cite{PhysRevX.15.021036}, 
and spinor Bose--Einstein condensates~\cite{Semenoff:2006vv,Kawaguchi:2012ii}.
In spinor condensates, noncommuting vortex charges lead to
nontrivial collision, braiding, and fusion
processes~\cite{Kobayashi:2008pk,Borgh:2016cco,Mawson:2018klj}.
Non-Abelian half-quantum vortices have also been investigated
in neutron $^3P_2$ superfluids~\cite{Masuda:2016vak,Masaki:2021hmk,Masaki:2023rtn},
where singly quantized vortices can split into pairs of
half-quantum vortices and form vortex
molecules~\cite{Kobayashi:2022moc,Kobayashi:2022dae}.
Related bound composite vortices have also been studied
in the $^1S_0$--$^3P_2$ coexisting
phase~\cite{Hattori:2026bvb}.
In gauge theories, a corresponding realization occurs
when a non-Abelian gauge group is Higgsed to a non-Abelian discrete subgroup, so that several distinct string types with noncommuting topological charges can arise \cite{Bais:1980vd}.

Gauge theories Higgsed to discrete subgroups can retain
observable Aharonov--Bohm interactions~\cite{Krauss:1988zc,Alford:1989ch,Preskill:1990bm}.
The fusion and braiding of excitations in non-Abelian discrete
gauge theories have also been studied in $2+1$
dimensions~\cite{Bais:1991pe}, including extensions with
Chern--Simons terms~\cite{Bais:1992ca}.
Finite-group gauge theories also play an important role in
anyonic quantum computation in $2+1$ dimensions
\cite{Mochon:2003zz,Mochon:2004hoi}
and in the study of generalized symmetries
\cite{Gaiotto:2014kfa},
including non-invertible symmetries
\cite{Heidenreich:2021xpr}.
Non-invertible symmetries can also impose selection rules
on topological soliton configurations, as demonstrated for
axionic domain walls and magnetic strings in axion
electrodynamics~\cite{Hidaka:2024kfx}. 
Topological descriptions of finite-group gauge theories capture flux, fusion, and braiding
data, but do not determine the core structure or energetic stability
of a particular vortex configuration.

In our previous studies~\cite{Abe:2026zrk,Abe:2026grk}, we studied one such example, an \(SO(3)\) gauge--Higgs theory in which \(SO(3)\) is Higgsed to the tetrahedral group \(A_4\) 
\cite{Ovrut:1977cn,Etesi:1996urw}.
Two types of axisymmetric finite-tension vortices were constructed. 
To our knowledge, Ref.~\cite{Abe:2026zrk} was the first to construct smooth
vortices with noncommuting topological charges and finite tension
in a renormalizable gauge--Higgs theory.
They are distinguished by their asymptotic \(SO(3)\) holonomy:
the two types are associated with order-two and order-three elements of \(A_4\), respectively, and are therefore referred to as the order-two and order-three vortices.
Correspondingly, they carry non-Abelian topological charges with nontrivial fusion rules \cite{Abe:2026zrk}. 
Both vortex families admit critically coupled, BPS-like solutions.
The static energies of these vortices were studied in Ref.~\cite{Abe:2026zrk}, and their linear stability was analyzed in Ref.~\cite{Abe:2026grk} 
by extending the analysis for the Abelian-Higgs vortices
\cite{GustafsonSigal2000}.
The stability depends on the Higgs and gauge-boson mass ratios, and the order-two vortex is linearly unstable in part of the parameter space.
The smooth, finite-tension solutions constructed in
Ref.~\cite{Abe:2026zrk}
provide an ultraviolet realization of the Gukov--Witten surface
operators \cite{Gukov:2008sn}
of the infrared $A_4$ gauge theory within a renormalizable
gauge--Higgs model.
This motivates us to investigate how unstable vortices
relax into separated constituents or bound molecules,
as a first step toward understanding non-Abelian string networks.

The first question addressed in the present work is how the
unstable order-two vortex relaxes.
We perturb the order-two vortex along a negative mode and follow
the gradient flow of the static energy.
We find that the vortex splits into two order-three vortices of opposite orientation. 
This fission process provides an explicit realization of a charge decomposition consistent with the non-Abelian fusion rules.

The second question concerns a possible finite-separation bound state of two order-three vortices.
In Ref.~\cite{Abe:2026grk}, we compared their binding energy with the linear stability of the axisymmetric vortex having the boundary condition corresponding to their combined topological charge.
Here, the binding energy measures the energy difference between this axisymmetric configuration and two infinitely separated order-three vortices.
We found a region of parameter space in which the axisymmetric vortex is linearly unstable while its energy is lower than that of the two infinitely separated vortices.
This indicated the possibility of a bound configuration in which the two order-three vortices remain at a finite separation.
In the present work, we verify this possibility numerically and find that such a finite-separation vortex molecule exists and is linearly stable, 
thus forming a molecule of non-Abelian vortices.

The remainder of this paper is organized as follows.
In Sec.~\ref{sec:model-nonlinear-setup}, we review the \(SO(3)\to A_4\) model, the two vortex types, and the linear stability results relevant to the present work, and introduce the gradient-flow framework used in the nonlinear analysis.
In Sec.~\ref{sec:order-two-decay}, we study the splitting of the unstable order-two vortex and identify the topological charges of the resulting two vortices.
In Sec.~\ref{sec:order-three-molecule}, we construct the order-three vortex molecule and analyze its linear stability.
Finally, Sec.~\ref{sec:summary-discussion} summarizes our results and discusses their implications.

\section{Model and preliminaries}
\label{sec:model-nonlinear-setup}
In this section, we summarize the ingredients needed for the nonlinear analysis.
We first review the model and the two vortex families studied in Ref.~\cite{Abe:2026zrk}.
We then recall the linear stability results relevant to the present work.
Finally, we introduce the gradient flow used to follow these instabilities beyond the linear regime and define the gauge-invariant vortex separation used in the subsequent analysis.

\subsection{Model and vortex families}

We consider the \(SO(3)\) gauge--Higgs model introduced in Ref.~\cite{Abe:2026zrk}, in which a real Higgs field transforms in the seven-dimensional, i.e., spin-three irreducible representation of \(SO(3)\).
The action is
\begin{align}
    S = \int d^4x\, \left[
        - \frac{1}{4e^2}G_{\mu\nu}^a G^{a\mu\nu}
        - \frac{1}{2}\langle \nabla_\mu \phi, \nabla^\mu \phi\rangle_{\bf 7}
        - V(\phi)
    \right].
\end{align}
Here $\langle\cdot, \cdot\rangle_{\ip}$ is the $SO(3)$-invariant inner product on the spin-three representation given below in Eq.~\eqref{eq:inner-product-7rep}.
$e>0$ is the gauge coupling, 
\begin{align}
    W_\mu = W_\mu^a T_a,
\end{align}
is the \(SO(3)\) gauge field, and
\begin{align}
    G_{\mu\nu}^a = \partial_\mu W_\nu^a - \partial_\nu W_\mu^a - \varepsilon^{abc} W_\mu^b W_\nu^c
\end{align}
is its field strength.
We choose the Hermitian generators \(T_a\) such that
\begin{align}
    [T_a,T_b] = \ii \varepsilon^{abc}T_c, \qquad \varepsilon^{123} = 1.
\end{align}
Explicitly, we take
\begin{align}
    \ii T_1 &=
    \begin{pmatrix}
        0 & 0 & 0\\
        0 & 0 & 1\\
        0 & -1 & 0
    \end{pmatrix}, \qquad
    \ii T_2 =
    \begin{pmatrix}
        0 & 0 & -1\\
        0 & 0 & 0\\
        1 & 0 & 0
    \end{pmatrix}, \qquad
    \ii T_3 =
    \begin{pmatrix}
        0 & 1 & 0\\
        -1 & 0 & 0\\
        0 & 0 & 0
    \end{pmatrix}.
    \label{eq:SO(3)-generator}
\end{align}
Let \(\rho\) denote the spin-three representation of \(SO(3)\).
For an infinitesimal gauge transformation
\begin{align}
    g(x) = \exp\!\left[\alpha^a(x)\ii T_a\right]
    = 1 + \alpha^a(x)\ii T_a + \mathcal O(\alpha^2),
\end{align}
we define the real representation matrices \(X_a\) for the spin-three representation by
\begin{align}
    \rho(g)\phi = \phi + \alpha^a X_a\phi + \mathcal O(\alpha^2),
    \qquad
    X_a\phi := \left.\frac{d}{d\alpha}\rho\!\left(e^{\alpha \ii T_a}\right)\phi\right|_{\alpha=0}.
\end{align}
With this convention,
\begin{align}
    [X_a,X_b] = -\varepsilon^{abc}X_c,
\end{align}
and the covariant derivative of \(\phi\) is
\begin{align}
    \nabla_\mu\phi = \partial_\mu \phi + W_\mu^a X_a \phi .
\end{align}

A convenient realization of the spin-three representation space is the space of real homogeneous harmonic polynomials of degree three,
\begin{align}
    \mathcal H_3 := \left\{
        f\in\mathbb R[\mathcal X,\mathcal Y,\mathcal Z]
        \,\middle|\,
        \deg f=3, \quad \Delta f=0
    \right\},
\end{align}
with the \(SO(3)\)-invariant inner product
\begin{align}
    \langle p,q\rangle_{\bf 7} := \frac{1}{4\pi}\int_{\mathcal S^2} p(\boldsymbol{\mathcal X}) q(\boldsymbol{\mathcal X})\,d\Omega .
    \label{eq:inner-product-7rep}
\end{align}
Here
\begin{align}
    \mathcal S^2 := \left\{
        (\mathcal X,\mathcal Y,\mathcal Z)\in\mathbb R^3
        \,\middle|\,
        \mathcal X^2+\mathcal Y^2+\mathcal Z^2=1
    \right\}.
\end{align}

\begin{figure}[t]
    \centering
    \includegraphics[width=1\textwidth]{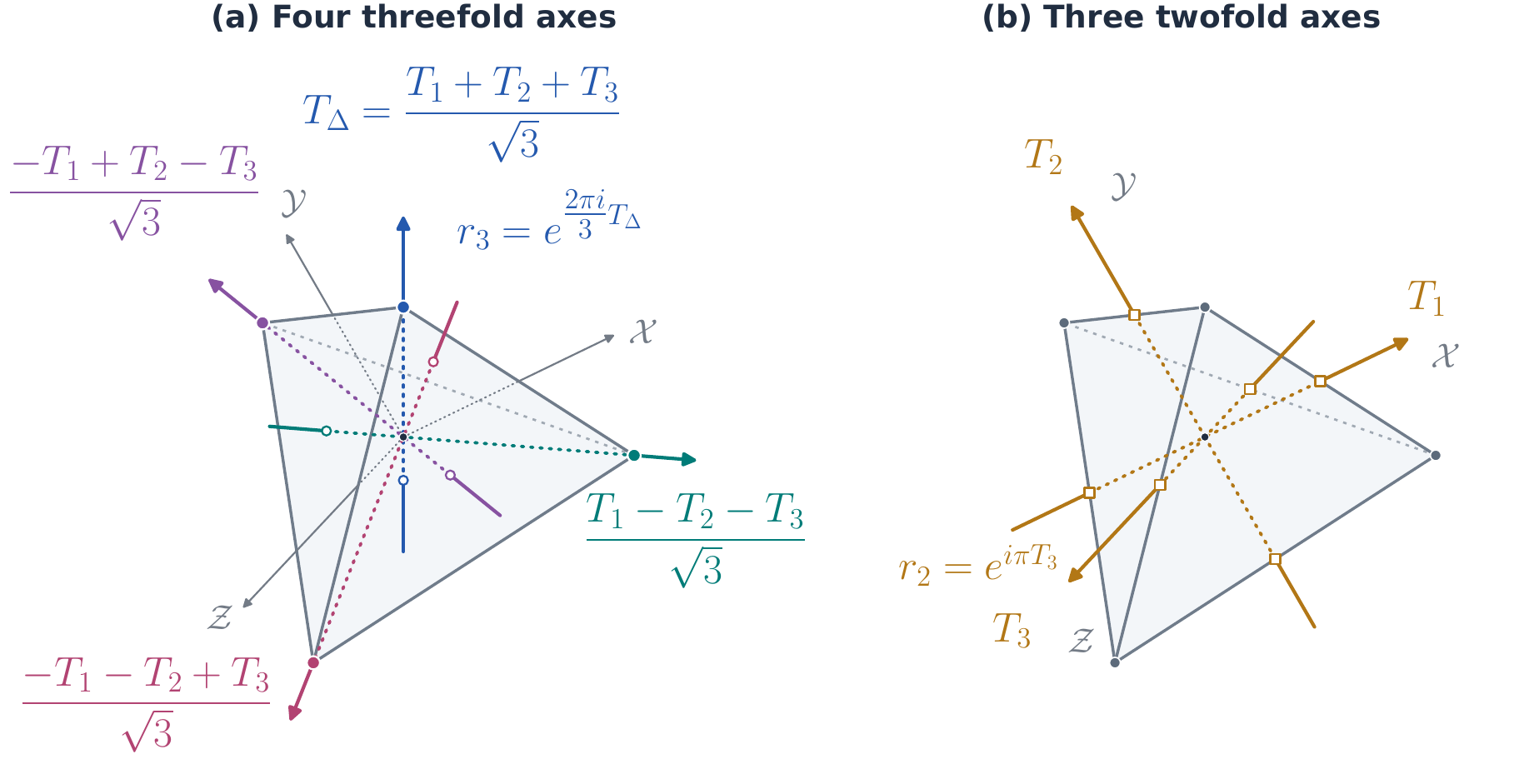}
    \caption{
Rotation axes of the tetrahedral group $A_4$ in the internal
$({\cal X},{\cal Y},{\cal Z})$ space.
(a) The four threefold axes pass through a vertex and the center
of the opposite face.
(b) The three twofold axes pass through the midpoints of opposite edges.
The corresponding $SO(3)$ generators are indicated beside the axes.
The group elements $r_3=\exp(2\pi i T_\Delta/3)$ and $r_2=\exp(i\pi T_3)$,
chosen in Eq.~(\ref{eq:A4-generators}), are also indicated beside their respective axes.
Coloured arrows point along the coefficient vectors of the displayed generators.
Axis segments inside the tetrahedron are dotted, while those outside are solid.
Filled circles mark vertices, open circles face centers,
and open squares edge midpoints.
The grey arrows in (a) indicate the coordinate axes,
which coincide with the twofold axes in (b).
}
    \label{fig:A4_rotation_axes}
\end{figure}

Throughout this paper we use an orthonormal basis \(\{f_A\}_{A=1}^7\),
\begin{align}
    f_1 &:= \frac{\sqrt{21}}{9}\Bigl(
        - 2\mathcal X^3
        + 3\mathcal X^2\mathcal Y
        + 3\mathcal X^2\mathcal Z
        + 3\mathcal X\mathcal Y^2
        + 15\mathcal X\mathcal Y\mathcal Z
        + 3\mathcal X\mathcal Z^2 \notag
    \\ &\hspace{27mm}
        - 2\mathcal Y^3
        + 3\mathcal Y^2\mathcal Z
        + 3\mathcal Y\mathcal Z^2
        - 2\mathcal Z^3
    \Bigr), 
     \label{eq:spin3basis0}
    \\[1mm]
    f_2 &:= \frac{\sqrt{7}}{6}\Bigl(
        - \mathcal X^3
        - 6\mathcal X^2\mathcal Y
        - 3\mathcal X^2\mathcal Z
        - 6\mathcal X\mathcal Y^2
        + 9\mathcal X\mathcal Z^2
        - \mathcal Y^3
        - 3\mathcal Y^2\mathcal Z
        + 9\mathcal Y\mathcal Z^2
        + 2\mathcal Z^3
    \Bigr), \\[1mm]
    f_3 &:= \frac{\sqrt{21}}{6}\Bigl(
        \mathcal X^3
        + 4\mathcal X^2\mathcal Y
        + 5\mathcal X^2\mathcal Z
        - 4\mathcal X\mathcal Y^2
        + \mathcal X\mathcal Z^2
        - \mathcal Y^3
        - 5\mathcal Y^2\mathcal Z
        - \mathcal Y\mathcal Z^2
    \Bigr), \\[1mm]
    f_4 &:= \frac{\sqrt{35}}{6} \Bigl(
        \mathcal X^3
        - 3\mathcal X^2\mathcal Y
        + 3\mathcal X^2\mathcal Z
        - 3\mathcal X\mathcal Y^2
        + \mathcal Y^3
        + 3\mathcal Y^2\mathcal Z
        - 2\mathcal Z^3
    \Bigr), \\[1mm]
    f_5 &:= \frac{\sqrt{105}}{6} \Bigl(
        \mathcal X^3
        + \mathcal X^2\mathcal Y
        - \mathcal X^2\mathcal Z
        - \mathcal X\mathcal Y^2
        - 2\mathcal X\mathcal Z^2
        - \mathcal Y^3
        + \mathcal Y^2\mathcal Z
        + 2\mathcal Y\mathcal Z^2
    \Bigr), \\[1mm]
    f_6 &:= \frac{\sqrt{105}}{18} \Bigl(
        2\mathcal X^3
        - 3\mathcal X^2\mathcal Y
        - 3\mathcal X^2\mathcal Z
        - 3\mathcal X\mathcal Y^2
        + 12\mathcal X\mathcal Y\mathcal Z
        - 3\mathcal X\mathcal Z^2 \notag
    \\ &\hspace{27mm}
        + 2\mathcal Y^3
        - 3\mathcal Y^2\mathcal Z
        - 3\mathcal Y\mathcal Z^2
        + 2\mathcal Z^3
    \Bigr), \\[1mm]
    f_7 &:= \frac{\sqrt{35}}{2} \Bigl(
        \mathcal X^2\mathcal Y
        - \mathcal X^2\mathcal Z
        - \mathcal X\mathcal Y^2
        + \mathcal X\mathcal Z^2
        + \mathcal Y^2\mathcal Z
        - \mathcal Y\mathcal Z^2
    \Bigr).
    \label{eq:spin3basis}
\end{align}
These basis vectors satisfy
\begin{align}
    \langle f_A,f_B\rangle_{\bf 7} = \delta_{AB}.
\end{align}
Accordingly, the Higgs field is expanded as
\begin{align}
    \phi = \sum_{A=1}^7 \phi_A f_A.
\end{align}
The Higgs potential is constructed from the quadratic invariant
\begin{align}
    I_2(\phi) := \langle\phi, \phi\rangle_{\bf 7}
\end{align}
and the non-negative quartic invariant
\begin{align}
    Q(\phi) := \frac{143}{315}I_4(\phi) - I_2(\phi)^2, \qquad
    I_4(\phi) := \frac{1}{4\pi}\int_{\mathcal S^2}\phi(\boldsymbol{\mathcal X})^4\,d\Omega .
\end{align}
We take the potential to be 
\begin{align}
    V(\phi) = \gamma Q(\phi) + \lambda\left(\langle\phi,\phi\rangle_{\bf 7} - v^2\right)^2,
\end{align}
with \(v^2,\lambda,\gamma>0\).
The vacuum conditions 
\begin{align}
    \langle\phi,\phi\rangle_{\bf 7} = v^2, \qquad Q(\phi) = 0
\end{align}
can be solved as  
\begin{align}
    \phi_{\rm vac} = \sqrt{105} v \mathcal{XYZ}
\end{align}
up to an \(SO(3)\) gauge transformation.
This vacuum is invariant under the tetrahedral subgroup \(A_4\subset SO(3)\).
A convenient choice of generators of \(A_4\) is
\begin{align}
    r_2 &=
    \begin{pmatrix}
        -1 & 0 & 0\\
        0 & -1 & 0\\
        0 & 0 & 1
    \end{pmatrix}, \qquad
    r_3 =
    \begin{pmatrix}
        0 & 1 & 0\\
        0 & 0 & 1\\
        1 & 0 & 0
    \end{pmatrix},
    \label{eq:A4-generators}
\end{align}
where \(r_2\) is a rotation by \(\pi\) around the \(\mathcal Z\)-axis and \(r_3\) is a rotation by \(2\pi/3\) around the \((1,1,1)\)-axis (see Fig.~\ref{fig:A4_rotation_axes}).
In particular, they satisfy 
\begin{align}
    r_2^2 = \boldsymbol 1_3, \qquad
    r_3^3 = \boldsymbol 1_3.
\end{align}

Expanding the theory around the vacuum gives four physical Higgs modes:
one of mass \(m_+\) and three degenerate modes of mass \(m_-\).
The remaining three Higgs components are absorbed by the three gauge bosons, which acquire a common mass \(m_G\), with
\begin{align}
    m_+ = 2\sqrt{2\lambda}v, \qquad
    m_- = \frac{4\sqrt{2\gamma}}{3\sqrt{5}}v, \qquad
    m_G = 2ev .
\end{align}
It is convenient to parametrize the theory by the two dimensionless mass ratios
\begin{align}
    \kappa := \frac{m_+}{m_G} = \frac{\sqrt{2\lambda}}{e}, \qquad
    \eta := \frac{m_+}{m_-} = \frac{3\sqrt{5}}{2}\sqrt{\frac{\lambda}{\gamma}}.
\end{align}

We introduce dimensionless variables
\begin{align}
    x^i \longrightarrow \tilde{x}^i = ev\,x^i, \qquad
    \phi \longrightarrow \tilde{\phi} = \frac{\phi}{v}, \qquad
    W_i^a \longrightarrow \tilde{W}^a_i = \frac{W_i^a}{ev},
    \label{eq:dimensionless}
\end{align}
and measure the dimensionful quantities in units of $v$.
We suppress the tildes below.
The tension, that is the energy per unit length of a static straight vortex then becomes
\begin{align}
    E[W,\phi] = \int_{\mathbb R^2} d^2x \left[
        \frac{1}{4}G_{ij}^a G_{ij}^a
        + \frac{1}{2}\langle \nabla_i\phi,\nabla_i\phi\rangle_{\bf 7}
        + U(\phi)
    \right],
    \label{eq:dimensionless-energy}
\end{align}
where
\begin{align}
    U(\phi) = \frac{45\kappa^2}{8\eta^2}Q(\phi)
    + \frac{\kappa^2}{2}\left(\langle\phi,\phi\rangle_{\bf 7} - 1\right)^2 .
\end{align}
The static Euler--Lagrange equations are
\begin{align}
    0 &= -\bigl(\nabla_i G_{ij}\bigr)^a + \langle X_a\phi, \nabla_j\phi\rangle_{\bf 7}, \\
    0 &= -\nabla_i\nabla_i\phi + \frac{45\kappa^2}{8\eta^2}\operatorname{grad}_{\phi}Q(\phi)
    + 2\kappa^2\bigl(\langle\phi,\phi\rangle_{\bf 7} - 1\bigr)\phi .
\end{align}
Here, \(\operatorname{grad}_{\phi} Q(\phi)\) denotes the gradient of \(Q\) with respect to the inner product \(\langle\cdot,\cdot\rangle_{\bf 7}\):
\begin{align}
    \left.\frac{d}{ds}Q(\phi+s\xi)\right|_{s=0}
    = \langle\operatorname{grad}_{\phi} Q(\phi),\xi\rangle_{\bf 7}
    \qquad \text{ for all } \xi \in \mathcal{H}_3.
\end{align}
For an adjoint-valued field \(\alpha = \alpha^aT_a\), we use
\begin{align}
    \bigl(\nabla_i\alpha\bigr)^a = \partial_i\alpha^a - \varepsilon^{abc}W_i^b\alpha^c .
\end{align}

\subsection{Vortex ans\"atze and topological charges}

Reference~\cite{Abe:2026zrk} constructed two families of axisymmetric finite-tension vortices.
They are characterized by the \(SO(3)\) holonomy around the vortex.
For a closed contour \(C\), we denote the holonomy by
\begin{align}
    \omega(C) := {\rm P}\exp\left( -\ii\oint_C W^a T_a \right),
    \label{eq:SO3-holonomy}
\end{align}
where the symbol ``P'' denotes the path-ordered product,
and $W^a= W^a_\mu dx^\mu$.
For a sufficiently large loop linking the vortex once, finite tension requires \(\omega(C)\) to approach an element of the unbroken subgroup \(A_4\).
We refer to vortices whose asymptotic holonomies are powers of the elements \(r_2\) and \(r_3\) defined in
Eq.~\eqref{eq:A4-generators} as the order-two and order-three families, respectively.

We first consider the order-two family.
It is convenient to introduce the two orthonormal vectors
\begin{align}
    \widehat v_c &:= \sqrt{105}\,\mathcal X\mathcal Y\mathcal Z, \\
    \widehat v_s &:= \frac{\sqrt{105}}{2} \mathcal Z\left(\mathcal X^2-\mathcal Y^2\right).
\end{align}
For each nonzero integer \(N\), the axisymmetric ansatz is
\begin{align}
    W &= -\frac{N}{2}K(r)T_3\,d\theta, \\
    \phi &= h(r) \left[\sin(N\theta)\widehat v_s + \cos(N\theta)\widehat v_c\right].
\end{align}
Regularity at the vortex core and finite tension require
\begin{align}
    K(0)&=0, \qquad
    h(0)=0, \\
    K(r)&\longrightarrow1, \qquad
    h(r)\longrightarrow1 \quad
    (r\longrightarrow\infty).
\end{align}
The holonomy along a sufficiently large loop \(C_\infty\) linking the vortex is then
\begin{align}
    \omega_\infty^{(2)}
    = \mathrm{P}\exp\left(-\ii\oint_{C_\infty} W\right)
    = \exp\left(\ii N\pi T_3\right)
    = r_2^N,
    \label{eq:order-two-holonomy}
\end{align}
where we fix the base point of the holonomy to $(r,\theta)= (\infty , 0) $.

We next consider the order-three family.
We introduce
\begin{align}
    T_\Delta := \frac{T_1+T_2+T_3}{\sqrt{3}}, \qquad
    X_\Delta := \frac{X_1+X_2+X_3}{\sqrt{3}},
\end{align}
which generate rotations about the \((1,1,1)\)-axis.
The orthonormal basis \eqref{eq:spin3basis0}--\eqref{eq:spin3basis} satisfies
\begin{align}
    X_\Delta f_1 &= 0, \nonumber\\
    X_\Delta f_{2q} &= -q f_{2q+1}, \qquad
    X_\Delta f_{2q+1} = q f_{2q}, \qquad
    q = 1, 2, 3.
    \label{eq:order-three-adapted-action}
\end{align}
These relations can be verified directly from the explicit polynomials in Eqs.~\eqref{eq:spin3basis0}--\eqref{eq:spin3basis} using the angular-momentum operators
\begin{align}
    X_1=\mathcal Y\partial_{\mathcal Z} - \mathcal Z\partial_{\mathcal Y},\quad
    X_2=\mathcal Z\partial_{\mathcal X} - \mathcal X\partial_{\mathcal Z},\quad
    X_3=\mathcal X\partial_{\mathcal Y} - \mathcal Y\partial_{\mathcal X}.
\end{align}

For each nonzero integer \(M\), the axisymmetric ansatz is
\begin{align}
    W &= -\frac{M}{3}J(r)T_\Delta\,d\theta, \label{eq:order-three-ansatz-W} \\
    \phi &= \frac{\sqrt{5}}{3}b(r)f_1 + \frac{2}{3}a(r)\left[\cos(M\theta)f_6 - \sin(M\theta)f_7\right].
    \label{eq:order-three-ansatz-phi}
\end{align}
Regularity at the vortex core requires
\begin{align}
    J(0)=0, \qquad
    a(0)=0, \qquad
    b'(0)=0,
\end{align}
while finite tension requires
\begin{align}
    J(r)\longrightarrow1, \qquad
    a(r)\longrightarrow1, \qquad
    b(r)\longrightarrow1 \quad
    (r\longrightarrow\infty).
    \label{eq:order-three-boundary}
\end{align}
The holonomy along a sufficiently large loop \(C_\infty\) linking the vortex is
\begin{align}
    \omega_\infty^{(3)}
    = \mathrm{P}\exp\left(-\ii\oint W\right)
    = \exp\left(\frac{2\pi\ii M}{3}T_\Delta\right)
    = r_3^M.
    \label{eq:order-three-holonomy}
\end{align}

We refer to \(N\) and \(M\) as the winding numbers of the order-two and order-three families, respectively.
Changing the sign of the winding number reverses the orientation of the vortex.
This terminology is particularly natural at the Bogomol'nyi-like mass ratios described below, where the vortex tensions are exactly proportional to the absolute values of these winding numbers.
We call the vortices with
\begin{align}
    |N|=1, \qquad |M|=1
\end{align}
the fundamental vortices of the corresponding families.

The \(SO(3)\) holonomy does not contain the full topological information.
Since the vacuum manifold is \(SO(3)/A_4\), the based vortex charge takes values in
\begin{align}
    \pi_1\!\left(SO(3)/A_4\right)
    \simeq \widetilde A_4
    := p^{-1}(A_4)
    \subset SU(2),
\end{align}
where
\begin{align}
    p : SU(2) \longrightarrow SO(3)
\end{align}
is the double covering map and \(\widetilde A_4\) is the binary tetrahedral group.

To obtain explicit expressions for the topological charges, we lift the \(SO(3)\) holonomies to \(SU(2)\).
We use the Hermitian \(SU(2)\) generators
\begin{align}
    \widetilde T_1 &:= \frac{1}{2}
    \begin{pmatrix}
        0 & 1\\
        1 & 0
    \end{pmatrix}, &
    \widetilde T_2 &:= \frac{1}{2}
    \begin{pmatrix}
        0 & -\ii\\
        \ii & 0
    \end{pmatrix}, &
    \widetilde T_3 &:= \frac{1}{2}
    \begin{pmatrix}
        1 & 0\\
        0 & -1
    \end{pmatrix}.
\end{align}
These are one half of the standard Pauli matrices and satisfy
\begin{align}
    [\widetilde T_a, \widetilde T_b] = \ii\varepsilon^{abc}\widetilde T_c,
\end{align}
with the same normalization as the \(SO(3)\) generators \(T_a\).
Accordingly, we choose the double covering map \(p:SU(2)\to SO(3)\) such that
\begin{align}
    p\!\left[\exp\!\left(\ii\omega^a\widetilde T_a\right)\right]
    = \exp\!\left(\ii\omega^aT_a\right).
\end{align}
For a closed contour \(C\), in addition to the \(SO(3)\) holonomy \(\omega(C)\) defined in Eq.~\eqref{eq:SO3-holonomy}, we define its \(SU(2)\) lift by
\begin{align}
    \widetilde\omega(C) := {\rm P}\exp\left(-\ii\oint_C W^a\widetilde T_a\right).
\end{align}
By construction,
\begin{align}
    p\!\left(\widetilde\omega(C)\right)=\omega(C).
\end{align}
For a sufficiently large loop \(C_\infty\) linking the vortex once,
\(\omega(C_\infty)\in A_4\), and hence
\begin{align}
    \widetilde\omega(C_\infty)
    \in p^{-1}(A_4)
    = \widetilde A_4.
\end{align}
The element \(\widetilde\omega(C_\infty)\) therefore represents the
based topological charge of the vortex.

For the order-two family, Eq.~\eqref{eq:order-two-holonomy} gives
\begin{align}
    \widetilde\omega_\infty^{(2)}
    &=
    \exp\left(\ii N\pi\widetilde T_3\right)
    =
    \widetilde r_2^{\,N},
\end{align}
where
\begin{align}
    \widetilde r_2
    &:=
    \exp\left(\ii\pi\widetilde T_3\right)
    =
    \begin{pmatrix}
        \ii & 0\\
        0 & -\ii
    \end{pmatrix}.
\end{align}

Similarly, for the order-three family, we define
\begin{align}
    \widetilde T_\Delta
    :=
    \frac{
        \widetilde T_1+\widetilde T_2+\widetilde T_3
    }{\sqrt{3}},
\end{align}
which is the \(SU(2)\) generator corresponding to the \(SO(3)\)
generator \(T_\Delta\).
Equation~\eqref{eq:order-three-holonomy} then gives
\begin{align}
    \widetilde\omega_\infty^{(3)}
    &=
    \exp\left(
        \frac{2\pi\ii M}{3}\widetilde T_\Delta
    \right)
    =
    \widetilde r_3^{\,M},
\end{align}
where
\begin{align}
    \widetilde r_3
    &:=
    \exp\left(
        \frac{2\pi\ii}{3}\widetilde T_\Delta
    \right)
    =
    \frac{1}{2}
    \begin{pmatrix}
        1+\ii & 1+\ii\\
        -1+\ii & 1-\ii
    \end{pmatrix}.
\end{align}

A gauge transformation, as well as a change of the base point of the loop, acts on the based topological charge by conjugation.
Therefore, the gauge-invariant topological charge is specified by a conjugacy class of \(\widetilde A_4\), which we denote by \(D_i\).

For later use, we also introduce the conjugation-invariant trace of the lifted holonomy in the spin-$1/2$ (2-dimensional) representation,
\begin{align}
    \widetilde\chi(C) := \Tr\widetilde\omega(C).
\end{align}
The topological classes realized by the two vortex families, together with the corresponding values of \(\widetilde\chi\), are summarized in Table~\ref{tab:theoretical-vortex-holonomies}.

\begin{table}[t]
    \centering
    \begin{tabular}{c|c|c|c|c}
        family      & winding        & topological class              & \(SO(3)\) holonomy class & \(\widetilde\chi\) \\
        \hline \hline
        order-two   & \(N=0\pmod 4\) & \([\boldsymbol 1_2]=D_0\)      & \([\boldsymbol 1_3]\)    & \(2\)  \\
        order-two   & \(N=1\pmod 4\) & \([\widetilde r_2]=D_6\)       & \([r_2]\)                & \(0\)  \\
        order-two   & \(N=2\pmod 4\) & \([-\boldsymbol 1_2]=D_1\)     & \([\boldsymbol 1_3]\)    & \(-2\) \\
        order-two   & \(N=3\pmod 4\) & \([\widetilde r_2^{-1}]=D_6\)  & \([r_2]\)                & \(0\)  \\
        \hline \hline
        order-three & \(M=0\pmod 6\) & \([\boldsymbol 1_2]=D_0\)      & \([\boldsymbol 1_3]\)    & \(2\)  \\
        order-three & \(M=1\pmod 6\) & \([\widetilde r_3]=D_2\)       & \([r_3]\)                & \(1\)  \\
        order-three & \(M=2\pmod 6\) & \([\widetilde r_3^{\,2}]=D_4\) & \([r_3^2]\)              & \(-1\) \\
        order-three & \(M=3\pmod 6\) & \([-\boldsymbol 1_2]=D_1\)     & \([\boldsymbol 1_3]\)    & \(-2\) \\
        order-three & \(M=4\pmod 6\) & \([\widetilde r_3^{\,4}]=D_5\) & \([r_3]\)                & \(-1\) \\
        order-three & \(M=5\pmod 6\) & \([\widetilde r_3^{-1}]=D_3\)  & \([r_3^2]\)              & \(1\)  \\
    \end{tabular}
    \caption{
        Gauge-invariant topological classes of the order-two and order-three vortex families and the corresponding lifted-holonomy traces \(\widetilde\chi=\Tr\widetilde\omega\).
        For \(g\in\widetilde A_4\) or \(g\in A_4\), \([g]\) denotes the conjugacy class containing \(g\).
    }
    \label{tab:theoretical-vortex-holonomies}
\end{table}

The lifted-holonomy trace does not distinguish a topological charge from its inverse.
Indeed, reversing the orientation of the vortex sends
\begin{align}
    \widetilde\omega \longrightarrow \widetilde\omega^{-1}.
\end{align}
Since \(\widetilde\omega\in SU(2)\), its trace is real, and hence
\begin{align}
    \Tr\widetilde\omega^{-1} = \overline{\Tr\widetilde\omega}
    = \Tr
\widetilde\omega .
\end{align}
Consequently, \(\widetilde\chi\) alone does not distinguish a vortex from the vortex with the opposite orientation.
For example, the \(M=1\) vortex carries the topological class \(D_2=[\widetilde r_3]\), whereas the \(M=-1\) vortex carries \(D_3=[\widetilde r_3^{-1}]\), but both have
\begin{align}
    \widetilde\chi = 1.
\end{align}

The possible fusion channels are determined by the multiplication of conjugacy classes in \(\widetilde A_4\).
For each conjugacy class \(D_i\), we define the corresponding  
formal class sum in the group algebra
by
\begin{align}
    \mathcal D_i := \sum_{g\in D_i} g,
\end{align}
whose gauge invariance is manifest, $g' \mathcal D_i g'^{-1} =\mathcal D_i$ for $g' \in \widetilde{A}_4$.
Their products can be written as
\begin{align}
    \mathcal D_i\mathcal D_j = \sum_k N_{ij}^{\phantom{ij}k}\mathcal D_k ,
\end{align}
where \(N_{ij}^{\phantom{ij}k}\) is the number of ordered pairs \((g_1,g_2)\in D_i\times D_j\) satisfying \(g_2 g_1 = g\) for a fixed element \(g\in D_k\).
This number is independent of the choice of \(g\) within \(D_k\).
The full multiplication table is shown in Table~\ref{tab:binary-tetrahedral-fusion}.

\begin{table}[t]
    \centering
    \resizebox{\textwidth}{!}{
    \begin{tabular}{c|ccccccc}
        & \(\mathcal D_0\)
        & \(\mathcal D_1\)
        & \(\mathcal D_2\)
        & \(\mathcal D_3\)
        & \(\mathcal D_4\)
        & \(\mathcal D_5\)
        & \(\mathcal D_6\)
        \\
        \hline
        \(\mathcal D_0\)
        & \(\mathcal D_0\)
        & \(\mathcal D_1\)
        & \(\mathcal D_2\)
        & \(\mathcal D_3\)
        & \(\mathcal D_4\)
        & \(\mathcal D_5\)
        & \(\mathcal D_6\)
        \\
        \(\mathcal D_1\)
        & \(\mathcal D_1\)
        & \(\mathcal D_0\)
        & \(\mathcal D_5\)
        & \(\mathcal D_4\)
        & \(\mathcal D_3\)
        & \(\mathcal D_2\)
        & \(\mathcal D_6\)
        \\
        \(\mathcal D_2\)
        & \(\mathcal D_2\)
        & \(\mathcal D_5\)
        & \(3\mathcal D_3+\mathcal D_4\)
        & \(4\mathcal D_0+2\mathcal D_6\)
        & \(4\mathcal D_1+2\mathcal D_6\)
        & \(\mathcal D_3+3\mathcal D_4\)
        & \(3\mathcal D_2+3\mathcal D_5\)
        \\
        \(\mathcal D_3\)
        & \(\mathcal D_3\)
        & \(\mathcal D_4\)
        & \(4\mathcal D_0+2\mathcal D_6\)
        & \(3\mathcal D_2+\mathcal D_5\)
        & \(\mathcal D_2+3\mathcal D_5\)
        & \(4\mathcal D_1+2\mathcal D_6\)
        & \(3\mathcal D_3+3\mathcal D_4\)
        \\
        \(\mathcal D_4\)
        & \(\mathcal D_4\)
        & \(\mathcal D_3\)
        & \(4\mathcal D_1+2\mathcal D_6\)
        & \(\mathcal D_2+3\mathcal D_5\)
        & \(3\mathcal D_2+\mathcal D_5\)
        & \(4\mathcal D_0+2\mathcal D_6\)
        & \(3\mathcal D_3+3\mathcal D_4\)
        \\
        \(\mathcal D_5\)
        & \(\mathcal D_5\)
        & \(\mathcal D_2\)
        & \(\mathcal D_3+3\mathcal D_4\)
        & \(4\mathcal D_1+2\mathcal D_6\)
        & \(4\mathcal D_0+2\mathcal D_6\)
        & \(3\mathcal D_3+\mathcal D_4\)
        & \(3\mathcal D_2+3\mathcal D_5\)
        \\
        \(\mathcal D_6\)
        & \(\mathcal D_6\)
        & \(\mathcal D_6\)
        & \(3\mathcal D_2+3\mathcal D_5\)
        & \(3\mathcal D_3+3\mathcal D_4\)
        & \(3\mathcal D_3+3\mathcal D_4\)
        & \(3\mathcal D_2+3\mathcal D_5\)
        & \(6\mathcal D_0+6\mathcal D_1+4\mathcal D_6\)
        \\
    \end{tabular}
    }
    \caption{
        Multiplication table of the conjugacy-class sums of the binary tetrahedral group \(\widetilde A_4\).
    }
    \label{tab:binary-tetrahedral-fusion}
\end{table}

Consider two vortices with based charges $g_1\in D_i$ and
$g_2\in D_j$, defined at a common base point using fixed
reference paths. With the reference vacuum fixed, gauge
transformations act on the two charges by simultaneous
conjugation,
\begin{align}
(g_1,g_2)\sim
(hg_1h^{-1},hg_2h^{-1}),
\qquad h\in\widetilde A_4.
\end{align}
Here $h$ is either $SU(2)$ lift of a residual $A_4$ gauge transformation at the common base point.
Since $h$ and $-h$ induce the same conjugation, this defines an action of $A_4\simeq\widetilde A_4/\{\pm\boldsymbol 1_2\}$
on the lifted charges.
The resulting equivalence class is called the simultaneous
conjugacy class of the ordered pair.
For fixed individual conjugacy classes $D_i$ and $D_j$,
we refer to the relative holonomy data encoded by this
simultaneous conjugacy class as the relative internal
orientation.
The total based charge is the ordered product $g_2g_1$.
Its conjugacy class is determined by the relative internal
orientation, although it need not determine that orientation
uniquely.
For example, for a pair with \(g_1\in D_2\) and \(g_2\in D_3\), one finds
\begin{align}
    \mathcal D_2 \mathcal D_3 = 4\mathcal D_0 + 2\mathcal D_6,
\end{align}
so that the fused configuration can belong to either \(D_0\) or \(D_6\), depending on the relative internal orientation.
Since \(|D_2|=|D_3|=4\), there are 16 ordered pairs of based charges.
Among them, \(4|D_0|=4\) give a product in \(D_0\), while \(2|D_6|=12\) give a product in \(D_6\).

As explicit realizations of these two fusion channels, we consider the following two configurations.
For the choice
\begin{align}
    g_1 = \widetilde r_3 \in D_2, \qquad
    g_2 = \widetilde r_3^{\;-1} \in D_3,
\end{align}
the total based charge is
\begin{align}
    g_2g_1 = \boldsymbol 1_2 \in D_0.
\end{align}
To obtain a different relative internal orientation, consider
\begin{align}
    \widetilde s_2 &:= \widetilde r_3^{\;-1}\widetilde r_2\widetilde r_3
    = \begin{pmatrix}
        0 & \ii\\
        \ii & 0
    \end{pmatrix}.
\end{align}
Conjugating \(\widetilde r_3^{\;-1}\) by \(\widetilde s_2\) gives another representative of \(D_3\),
\begin{align}
    g_2' &:= \widetilde s_2 \widetilde r_3^{\;-1} \widetilde s_2^{\;-1} \nonumber\\
    &= \frac{1}{2}
    \begin{pmatrix}
        1+\ii & 1-\ii\\
        -1-\ii & 1-\ii
    \end{pmatrix}
    \in D_3.\label{eq:g2'}
\end{align}
The total based charge is then
\begin{align}
    g_2'g_1 = \widetilde r_2 \in D_6.
    \label{eq:two-three}
\end{align}
The two configurations above have the same individual vortex
classes $D_2$ and $D_3$, but belong to different simultaneous
conjugacy classes, since their total charges lie in $D_0$
and $D_6$, respectively.
The replacement $g_2\to g_2'$ in Eq.~(\ref{eq:g2'}) therefore changes
the relative internal orientation and is not a gauge
transformation of the original two-vortex configuration.
The two configurations already belong to distinct total
topological sectors before fusion.

Having summarized the topological classification and fusion properties, we next recall an energetic property of the two vortex families that will provide useful reference values in the nonlinear analysis.
At special values of the mass ratios, their tensions can be evaluated exactly and are proportional to the absolute values of the corresponding winding numbers.
For the order-two family, the reduced equations of motion and energy are independent of \(\eta\), and at \(\kappa=1\) they reduce to those of the critically coupled Abelian-Higgs model \cite{Bogomolny:1975de,Jaffe:1980mj,Abe:2026zrk}.
In the normalization of Eq.~\eqref{eq:dimensionless-energy},
\begin{align}
    E_N^{(2)} = \pi |N|
    \qquad (\kappa=1).
    \label{eq:order-two-exact-tension}
\end{align}
For the order-three family, at \(\eta=1\) the profile function \(b(r)\) is fixed to \(1\), and at \(\kappa=1\) the remaining system again reduces to the critically coupled Abelian-Higgs model \cite{Abe:2026zrk}, giving
\begin{align}
    E_M^{(3)} = \frac{4\pi}{9} |M|
    \qquad (\kappa=\eta=1).
\end{align}
These exact tensions provide useful reference values for the analysis below.
We next review the linear stability properties of the two vortex families that are relevant to their nonlinear evolution.

\subsection{Review of the linear stability analysis}
\label{subsec:review-linear-stability}

In Ref.~\cite{Abe:2026grk}, we studied the linear energetic stability of the two vortex families introduced above 
by extending the analysis for the Abelian-Higgs vortices
\cite{GustafsonSigal2000}.
Here we briefly review the results of the linear stability analysis that are relevant to the nonlinear evolution studied below.
For a static vortex background, we write collectively
\begin{align}
    \Psi := (W_x, W_y, \phi) = (W_i, \phi),
\end{align}
and denote an infinitesimal fluctuation by
\begin{align}
    \delta\Psi := (B_i, \xi).
\end{align}
We consider perturbations of the background fields of the form
\begin{align}
    W_i^a &\longmapsto W_i^a + s B_i^a,\\
    \phi &\longmapsto \phi + s \xi ,
\end{align}
where \(s\) is a small real parameter.
Since the background satisfies the static field equations, the energy has the expansion
\begin{align}
    E[\Psi + s\delta\Psi] = E[\Psi] + \frac{s^2}{2} \langle\delta\Psi, L\delta\Psi\rangle + \mathcal O(s^3),
\end{align}
where \(L\) is the Hessian of the energy.
Here, the inner product between two fluctuations $\delta \Psi=(B_i, \xi)$ and $\delta\Psi'=(B_i', \xi')$ is defined by 
\begin{align}
    \langle \delta\Psi, \delta\Psi'\rangle
    = \int_{\mathbb{R}^2} d^2x \, \left(B_i^a B_i^{\prime a} + \langle \xi, \xi' \rangle_{\bf 7} \right).
    \label{eq:inner-product}
\end{align}

Because \(L\) contains an infinite-dimensional kernel associated with infinitesimal gauge transformations, the stability analysis was performed in the background gauge
\begin{align}
    \Gamma^a(B,\xi)
    := \bigl(-\nabla_iB_i\bigr)^a
    - \langle X_a\phi,\xi\rangle_{\bf 7}
    = 0.
    \label{eq:review-background-gauge}
\end{align}
Equivalently, one may use the gauge-fixed Hessian \(\widetilde L\), obtained by adding the non-negative gauge-fixing term
\begin{align}
    \left\langle (B,\xi),\widetilde L(B,\xi)\right\rangle
    :=
    \left\langle (B,\xi),L(B,\xi)\right\rangle
    + \int d^2x\,\Gamma^a(B,\xi)\Gamma^a(B,\xi).
\end{align}
The additional term vanishes on the background-gauge subspace and lifts the gauge zero modes without changing the physical negative or zero modes.
A negative eigenvalue of \(\widetilde L\) therefore identifies a physical direction along which the vortex energy decreases at quadratic order.
Here, linear energetic stability refers to the absence of physical negative modes of the static Hessian.

We first consider the order-two vortex at
\begin{align}
    N=1, \qquad \kappa=\eta=1.
\end{align}
Although the order-two vortex at \(\kappa=1\) has the same reduced equations of motion and tension as the critically coupled Abelian-Higgs vortex, Ref.~\cite{Abe:2026grk} showed that the \(N=1\) vortex is unstable against fluctuations in the full \(SO(3)\) gauge theory.
At \((\kappa,\eta)=(1,1)\), there are two linearly independent eigenmodes with negative eigenvalue.

To write these modes, we introduce the two orthonormal Higgs vectors
\begin{align}
    \widehat q_c
    &:= \frac{\sqrt{42}}{4}\mathcal X
    \left(\mathcal X^2+\mathcal Y^2-4\mathcal Z^2\right), \notag\\
    \widehat q_s
    &:= \frac{\sqrt{42}}{4}\mathcal Y
    \left(-\mathcal X^2-\mathcal Y^2+4\mathcal Z^2\right).
\end{align}
Taking the real and imaginary parts of the complex negative mode
constructed in Sec.~4.4 of Ref.~\cite{Abe:2026grk},
we obtain the following two real modes, up to an overall normalization:
\begin{align}
    \delta\Psi_-^{(1)}:\qquad
    B_x &= F(r)T_1, \qquad
    B_y = -F(r)T_2, \qquad
    \xi = \sqrt{2}F(r)\widehat q_c,
    \label{eq:review-order-two-negative-mode-1}\\
    \delta\Psi_-^{(2)}:\qquad
    B_x &= F(r)T_2, \qquad
    B_y = F(r)T_1, \qquad
    \xi = \sqrt{2}F(r)\widehat q_s,
    \label{eq:review-order-two-negative-mode-2}
\end{align}
where
\begin{align}
    F(r)
    := \sqrt{\frac{h(r)}{r}}\,
    \exp\left[-\sqrt{5}\int_0^r h(w)\,dw\right].
\end{align}
Here \(h(r)\) is the Higgs profile of the \(N=1\) order-two vortex at \(\kappa=1\).
Both modes satisfy
\begin{align}
    \widetilde L\,\delta\Psi_-^{(a)}
    = -\delta\Psi_-^{(a)},
    \qquad a=1,2.
\end{align}
In particular, the gauge-field fluctuations in these negative modes are along \(T_1\) and \(T_2\), which do not commute with the background gauge generator \(T_3\).
This reflects the genuinely non-Abelian nature of the instability.

Our numerical spectral analysis revealed no negative modes other than these two.
The two modes are related by the rotational symmetry of the axisymmetric vortex background and therefore represent the same instability with different orientations in the transverse plane.
For the nonlinear analysis, it is therefore sufficient to choose one of the two modes as the initial perturbation.

We next turn to the order-three \(M=2\) vortex.
In Ref.~\cite{Abe:2026grk}, we conjectured the existence of a finite-separation molecular state by comparing the binding energy of the \(M=2\) vortex with its linear instability.
The basic idea is that an axisymmetric \(M=2\) vortex may be locally unstable against separation even when its energy is lower than that of two infinitely separated \(M=1\) vortices.

To characterize these two properties, we define the binding-energy difference
\begin{align}
    \Delta_{\rm bin}(\kappa,\eta)
    := E_2^{(3)}(\kappa,\eta) - 2E_1^{(3)}(\kappa,\eta),
\end{align}
where \(E_2^{(3)}(\kappa,\eta)\) and \(E_1^{(3)}(\kappa,\eta)\)
denote the dimensionless tensions of the axisymmetric order-three
vortices with winding numbers \(M=2\) and \(M=1\), respectively.
We denote by
\begin{align}
    \lambda_{\min}^{(2)}(\kappa,\eta)
\end{align}
the lowest non-translational eigenvalue of the gauge-fixed Hessian around the axisymmetric \(M=2\) vortex.
Thus,
\begin{align}
    \Delta_{\rm bin}<0
\end{align}
means that the \(M=2\) vortex has lower energy than two infinitely separated \(M=1\) vortices, whereas
\begin{align}
    \lambda_{\min}^{(2)}<0
\end{align}
signals a local instability of the axisymmetric configuration.

\begin{figure}[t]
    \centering
    \includegraphics[width=0.8\textwidth]{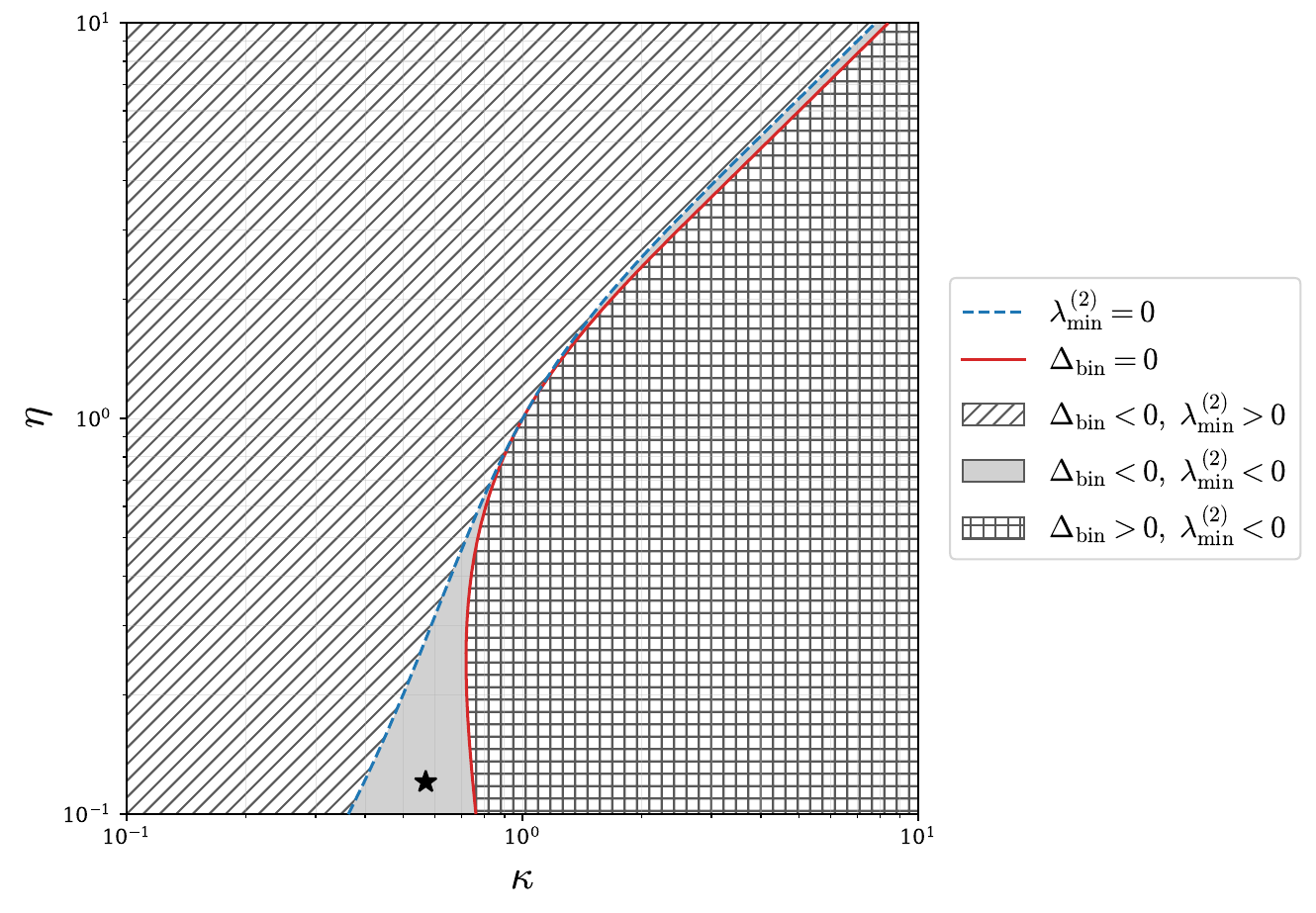}
    \caption{
        Zero contours of the binding-energy difference
        \(\Delta_{\rm bin}=E_2^{(3)}-2E_1^{(3)}\)
        and the lowest non-translational Hessian eigenvalue
        \(\lambda_{\min}^{(2)}\)
        for the order-three \(M=2\) vortex.
        The marked point indicates the parameter values
        \((\kappa_\star,\eta_\star)\)
        used in the nonlinear analysis of Sec.~\ref{sec:order-three-molecule}.
        It lies in the region satisfying
        \(\Delta_{\rm bin}<0\) and \(\lambda_{\min}^{(2)}<0\).
    }
    \label{fig:review-order-three-binding-instability}
\end{figure}

Figure~\ref{fig:review-order-three-binding-instability} shows the zero contours of these two quantities in the \((\kappa,\eta)\) plane.
The two contours do not coincide, leaving a finite region in which
\begin{align}
    \Delta_{\rm bin}<0, \qquad
    \lambda_{\min}^{(2)}<0.
\end{align}
In this region, an infinitesimal deformation along the unstable direction lowers the energy of the coincident \(M=2\) vortex, while complete separation into two isolated \(M=1\) vortices would increase the energy.
This observation led us in Ref.~\cite{Abe:2026grk} to conjecture that the instability may terminate at a finite vortex separation, producing a molecular state.

In the present work, we choose the parameter point
\begin{align}
    (\kappa_\star,\eta_\star)
    \simeq (0.56899,\,0.12068),
\end{align}
as indicated by $\star$ in Fig.~\ref{fig:review-order-three-binding-instability}.
At this point, the numerical stability analysis confirms the existence of two linearly independent negative modes.
We follow the notation of Ref.~\cite{Abe:2026grk} for the two negative modes below.
Up to an overall normalization, the first negative mode can be written as
\begin{align}
    \widehat{\delta\Psi}_-^{(1)}:\qquad
    B_x &= \left[
        S_+(r)\sin\theta + S_-(r)\sin(3\theta)
    \right]T_\Delta, \notag\\
    B_y &= \left[
        S_+(r)\cos\theta - S_-(r)\cos(3\theta)
    \right]T_\Delta,
    \label{eq:review-order-three-negative-gauge-mode-1}\\
    \xi &= \sqrt{2}\,\chi_0(r)\cos(2\theta)f_1 \notag\\
    &\quad
    + \left[
        \chi_{+3}(r)\cos(4\theta)-\chi_{-3}(r)
    \right]f_6
    - \chi_{+3}(r)\sin(4\theta)f_7 .
\end{align}
Here \(S_\pm(r)\), \(\chi_0(r)\), and \(\chi_{\pm3}(r)\) are real radial functions obtained numerically from the Hessian eigenvalue problem.
The second independent negative mode can be written as
\begin{align}
    \widehat{\delta\Psi}_-^{(2)}:\qquad
    B_x &= \left[
        S_+(r)\cos\theta + S_-(r)\cos(3\theta)
    \right]T_\Delta, \notag\\
    B_y &= \left[
        -S_+(r)\sin\theta + S_-(r)\sin(3\theta)
    \right]T_\Delta, \\
    \xi &= -\sqrt{2}\,\chi_0(r)\sin(2\theta)f_1
    - \chi_{+3}(r)\sin(4\theta)f_6 \notag\\
    &\quad
    - \left[
        \chi_{+3}(r)\cos(4\theta)+\chi_{-3}(r)
    \right]f_7 .
    \label{eq:review-order-three-negative-Higgs-mode-2}
\end{align}
The two modes are related by a spatial rotation and therefore describe the same instability with different orientations in the transverse plane.

An important feature of Eqs.~\eqref{eq:review-order-three-negative-gauge-mode-1}--\eqref{eq:review-order-three-negative-Higgs-mode-2} is that the gauge-field perturbations are proportional only to \(T_\Delta\), while the Higgs perturbations contain only the \(f_1\), \(f_6\), and \(f_7\) components.
These are precisely the field components appearing in the order-three vortex ansatz in Eqs.~\eqref{eq:order-three-ansatz-W} and \eqref{eq:order-three-ansatz-phi}.
Together with the energetic argument summarized in Fig.~\ref{fig:review-order-three-binding-instability}, this suggests that the unstable \(M=2\) vortex can evolve within the same field sector toward a finite-separation configuration of two \(M=1\) vortices.
The special structure of this field sector and its relation to an embedded Abelian-Higgs model will be discussed in more detail in Sec.~\ref{subsec:embedded-abelian-sector}.

To determine the nonlinear evolution of these instabilities, we
introduce the gradient flow in the next subsection.

\subsection{Gradient flow and vortex separation}
To determine the nonlinear configurations reached along the unstable directions reviewed above, we evolve \(\Psi\) by the gradient flow of the static energy,
\begin{align}
    \frac{\partial\Psi}{\partial\tau} = -\operatorname{grad}E[\Psi],
    \label{eq:gradient-flow}
\end{align}
where \(\tau\) denotes the flow time and \(\operatorname{grad}E\)
is the gradient of \(E\) with respect to the inner product
defined in Eq.~\eqref{eq:inner-product}. Explicitly,
\begin{align}
    \frac{\partial W_j^a}{\partial\tau}
    &= \bigl(\nabla_iG_{ij}\bigr)^a
    - \langle X_a\phi,\nabla_j\phi\rangle_{\bf 7},
    \label{eq:gradient-flow-gauge}\\
    \frac{\partial\phi}{\partial\tau}
    &= \nabla_i\nabla_i\phi
    - \frac{45\kappa^2}{8\eta^2}\operatorname{grad}_{\phi} Q(\phi)
    - 2\kappa^2\bigl(\langle\phi,\phi\rangle_{\bf 7}-1\bigr)\phi .
    \label{eq:gradient-flow-Higgs}
\end{align}
The right-hand sides of these equations vanish when the static field equations are satisfied.
Along the flow,
\begin{align}
    \frac{dE}{d\tau}
    = -\int d^2x\left[
        \sum_{i,a}\left(\frac{\partial W_i^a}{\partial\tau}\right)^2
        + \left\langle
            \frac{\partial\phi}{\partial\tau},
            \frac{\partial\phi}{\partial\tau}
        \right\rangle_{\bf 7}
    \right]
    \leq 0.
\end{align}
Thus, the gradient flow provides a direct probe of the energy landscape beyond the quadratic approximation of the Hessian analysis.
We emphasize that \(\tau\) is an auxiliary flow time and should not be interpreted as the physical time of the Lorentzian gauge--Higgs theory.

For an unstable vortex \(\Psi_\star\), we take the initial configuration to be
\begin{align}
    \Psi(\tau = 0)
    = \Psi_\star + \epsilon\,\delta\Psi_-,
    \label{eq:gradient-flow-initial-condition}
\end{align}
where \(\delta\Psi_-\) is one of the negative modes reviewed above and \(\epsilon>0\) is a small perturbation amplitude.

For the order-two vortex, the gradient flow is performed in the full field space of the six spatial gauge-field components and the seven Higgs components.
For the order-three \(M=2\) vortex, both the background and the negative mode lie in the five-component field sector
\begin{align}
    W_i = A_iT_\Delta, \qquad
    \phi = \phi_1f_1+\phi_6f_6+\phi_7f_7.
\end{align}
The order-three flow can therefore be studied within this sector.
Section~\ref{sec:order-three-molecule} shows that the full gradient flow preserves this sector and gives the corresponding Abelian Higgs model.

In the order-three analysis, we also perform an independent gradient-flow calculation starting from a configuration constructed from two initially separated \(M=1\) vortices.
This provides a complementary test of whether the unstable \(M=2\) vortex and a separated two-vortex configuration approach the same nonlinear configuration.

To characterize the splitting of a vortex, we define the vortex-core positions in a gauge-invariant manner.
At each point in the transverse plane, we introduce the symmetric matrix
\begin{align}
    \mathcal M_{ab}(x,y)
    := \langle X_a\phi(x,y),X_b\phi(x,y)\rangle_{\bf 7}
\end{align}
and denote its smallest eigenvalue by
\begin{align}
    \mu(x,y)
    := \lambda_{\min}\!\left[\mathcal M(x,y)\right].
    \label{eq:mu-core-diagnostic}
\end{align}
Since \(\mathcal M\) transforms by an orthogonal similarity transformation under a gauge transformation, its eigenvalues are gauge invariant.
For the vacuum,
\begin{align}
    \mathcal M_{ab}=4\delta_{ab},
    \qquad
    \mu=4,
\end{align}
whereas vortex cores appear as local minima of \(\mu\).

When two distinct cores are present at flow time \(\tau\), we identify their positions with the two local minima of \(\mu\),
\begin{align}
    \boldsymbol x_1(\tau),\qquad
    \boldsymbol x_2(\tau),
\end{align}
and define the vortex separation by
\begin{align}
    d_{\rm core}(\tau)
    := \left|
        \boldsymbol x_1(\tau)-\boldsymbol x_2(\tau)
    \right|.
\end{align}
This gauge-invariant core separation \(d_{\rm core}\), expressed in units of \((ev)^{-1}\) as in Eq.~\eqref{eq:dimensionless},
will be used below to characterize the nonlinear evolution of the vortex configurations.

\section{Decay of the order-two \texorpdfstring{\(N=1\)}{N=1} vortex}
\label{sec:order-two-decay}

\subsection{Preliminary}
We first investigate the nonlinear fate of the unstable order-two \(N=1\) vortex at
\begin{align}
    \kappa=\eta=1 .
\end{align}
As reviewed in Sec.~\ref{subsec:review-linear-stability}, the gauge-fixed Hessian evaluated at this vortex has a two-dimensional negative eigenspace with eigenvalue \(-1\).
The two modes given in Eqs.~\eqref{eq:review-order-two-negative-mode-1} and~\eqref{eq:review-order-two-negative-mode-2} are related by a spatial rotation.
We choose the mode \(\delta\Psi_-^{(1)}\) in Eq.~\eqref{eq:review-order-two-negative-mode-1} and use the initial condition \eqref{eq:gradient-flow-initial-condition}, with perturbation amplitude
\begin{align}
    \epsilon = 10^{-2} .
\end{align}

We use a square computational domain
\begin{align}
    -6 \leq x,y \leq 6 ,
\end{align}
with a uniform lattice spacing
\begin{align}
    \Delta x = \Delta y = 0.04 ,
\end{align}
corresponding to \(301\times301\) lattice points, and integrate the gradient flow with
\begin{align}
    \Delta\tau = 2.0\times10^{-4}.
\end{align}

The gradient flow of the weakly perturbed axisymmetric vortex shows that it splits into two vortices.
We first study the splitting process and then determine the topological charge carried by each separated vortex.

\subsection{Vortex splitting}

\begin{figure}[t]
    \centering
    \includegraphics[width=0.98\textwidth]{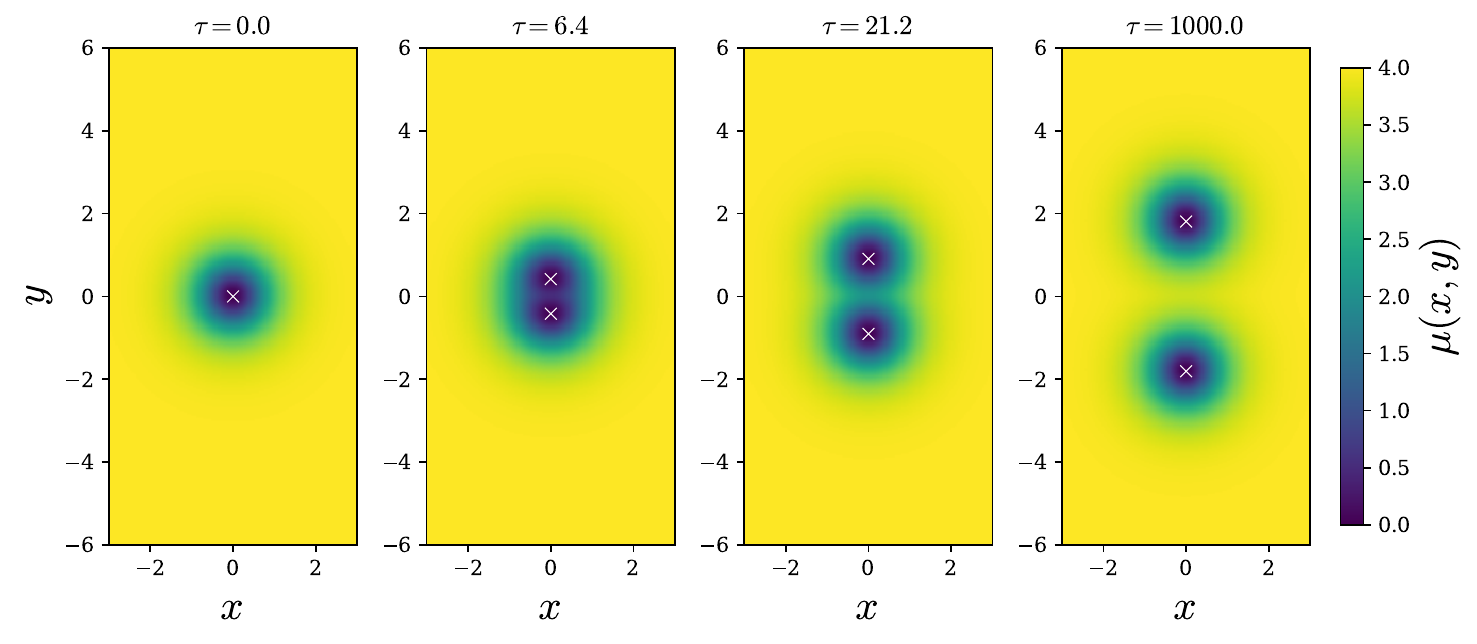}
    \caption{
        Splitting of the order-two \(N=1\) vortex under the gradient flow at \(\kappa=\eta=1\).
        The color map shows \(\mu(x,y)\), and the white crosses mark the vortex-core positions.
        The four panels show, from left to right, the initial state of the flow, the stage at which two distinct cores first become visible, and two representative configurations at later flow times.
        The corresponding flow time \(\tau\) is indicated above each panel.
    }
    \label{fig:order-two-decay-snapshots}
\end{figure}

We now follow the unstable order-two \(N=1\) vortex along the gradient flow.
The development of the instability is shown in Fig.~\ref{fig:order-two-decay-snapshots}, where we plot \(\mu(x,y)\) defined in Eq.~\eqref{eq:mu-core-diagnostic}.
The white crosses indicate the vortex-core positions identified from the
local minima of \(\mu\).

At \(\tau=0\), the configuration has a single core centered near the origin.
The negative-mode perturbation breaks the rotational symmetry of the axisymmetric vortex and grows under the flow.
By the flow time shown in the second panel of Fig.~\ref{fig:order-two-decay-snapshots}, two distinct cores have become visible in the \(\mu\) profile, and the subsequent snapshots show the splitting of the original vortex into two separated cores.

\begin{figure}[t]
    \centering
    \includegraphics[width=0.80\textwidth]{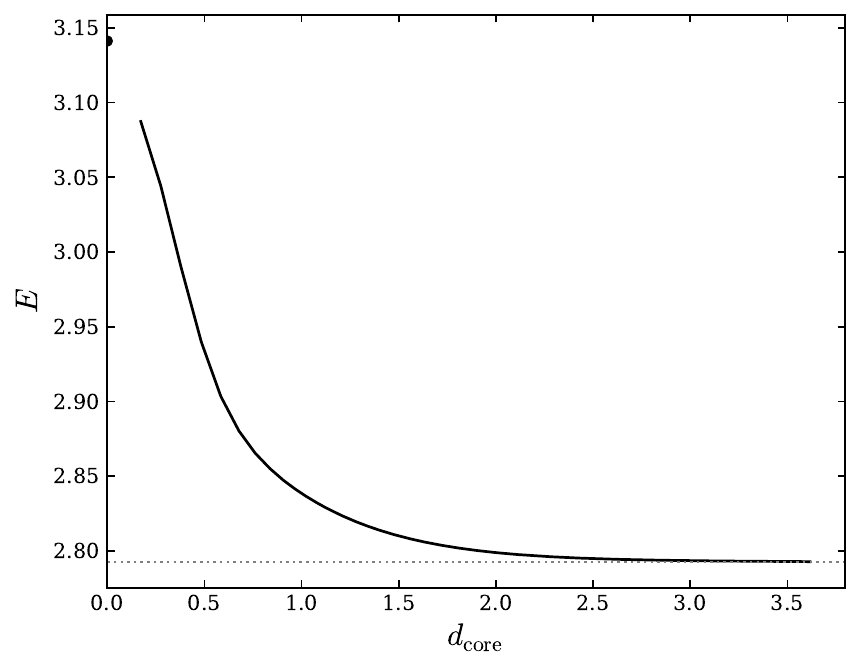}
    \caption{
        Total energy as a function of the vortex separation \(d_{\rm core}\) for the order-two \(N=1\) gradient flow at \(\kappa=\eta=1\).
The solid curve traces the gradient-flow trajectory
$(d_{\rm core}(\tau),E(\tau))$ after the two vortex cores
have become resolved.
It is not obtained by minimizing the energy separately
at each fixed vortex separation. 
        The point at \(d_{\rm core}=0\) marks the exact energy \(E_1^{(2)}=\pi\) of the unperturbed axisymmetric order-two vortex.
        The gray dotted line indicates the energy \(2E_1^{(3)}=8\pi/9\) of two isolated order-three \(M=1\) vortices.
    }
    \label{fig:order-two-energy-vs-distance}
\end{figure}

Figure~\ref{fig:order-two-energy-vs-distance} shows the total energy as a function of the vortex separation.
Since \(d_{\rm core}\) is defined only after two distinct minima of \(\mu\) have been resolved, only the resolved part of the gradient-flow trajectory (\(\tau\geq 6.4\); see Fig.~\ref{fig:order-two-decay-snapshots}) is connected by the solid curve.
For reference, the point at \(d_{\rm core}=0\) shows the exact energy \(E_1^{(2)}=\pi\) of the unperturbed axisymmetric order-two vortex, as given by Eq.~\eqref{eq:order-two-exact-tension}.
As we will show in the next subsection, the two vortices appearing in the late-time configuration are fundamental order-three vortices.
Accordingly, Fig.~\ref{fig:order-two-energy-vs-distance} also shows the reference energy \(2E_1^{(3)}\) of two isolated order-three \(M=1\) vortices.
The gradient-flow trajectory appears to approach this value as the vortex separation increases.

We next identify the topological charges of the two separated vortices directly from their Wilson loops.

\subsection{Topological identification of the daughter vortices}

\begin{figure}[t]
    \centering
    \includegraphics[width=0.68\textwidth]{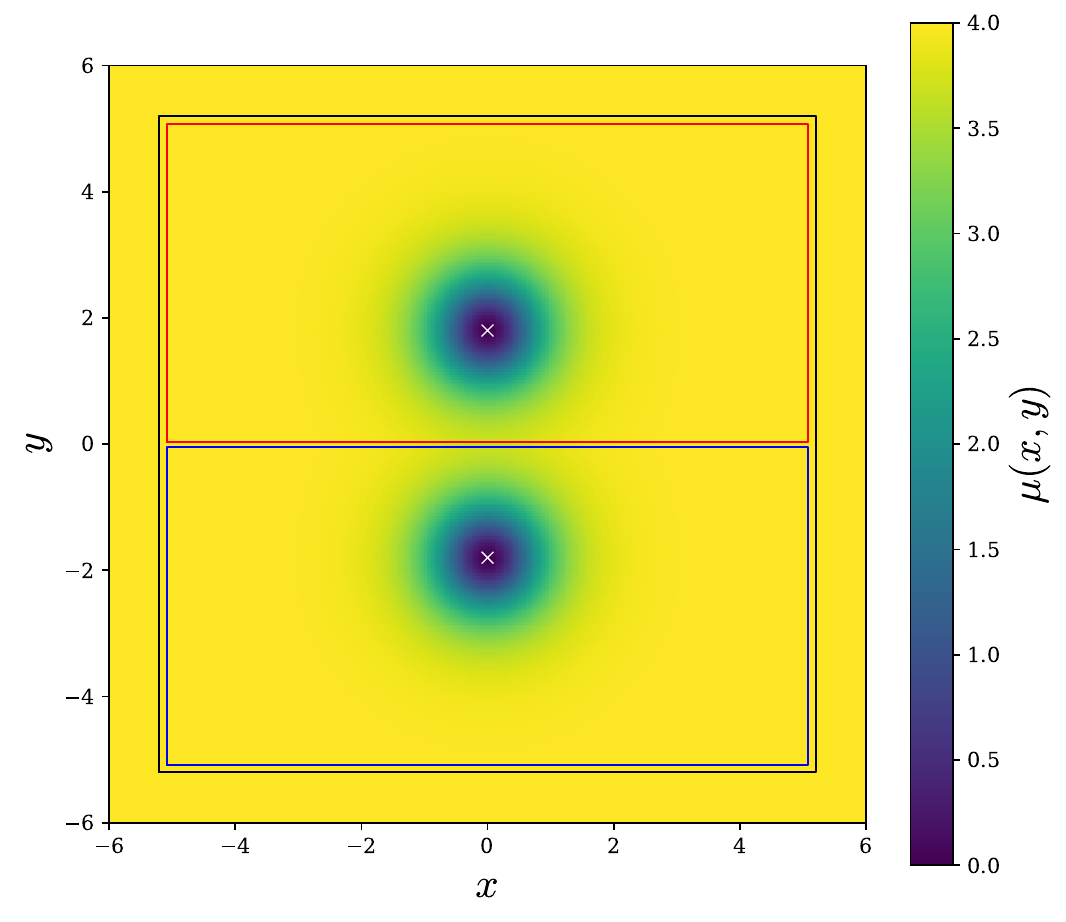}
    \caption{
        Contours used to evaluate \(\widetilde\chi(C)\) for the two separated vortices at \(\tau=1000\).
        The blue and red contours surround the lower and upper vortex cores, respectively, while the black contour encloses both cores.
        The color map shows \(\mu(x,y)\), and the white crosses mark the detected core positions.
    }
    \label{fig:order-two-daughter-wilson-loops}
\end{figure}

We now identify the topological charges carried by the two vortices resulting from the splitting of the order-two \(N=1\) vortex.
We evaluate the lifted-holonomy trace \(\widetilde\chi(C)\) along contours surrounding the separated vortex cores.
We use the late-time configuration at
\begin{align}
    \tau = 1000,
\end{align}
where the two cores are well separated (see Fig.~\ref{fig:order-two-decay-snapshots}).
The contours used in the calculation are shown in Fig.~\ref{fig:order-two-daughter-wilson-loops}.
The blue and red contours surround the lower and upper cores, respectively, while the black contour surrounds both vortices.
Each path starts at the lower-left corner and goes once counterclockwise around the rectangle.

The resulting values of \(\widetilde\chi\) are listed in Table~\ref{tab:order-two-daughter-wilson-results}.
In the exact theory, \(\widetilde\chi\) is real because \(\widetilde\omega\in SU(2)\).
The numerical values of \(\widetilde\chi\) acquire small imaginary parts from numerical errors, but these are much smaller than unity for all three contours and do not affect the identification of the topological sectors.

\begin{table}[t]
    \centering
    \begin{tabular}{c||c|c|c}
        contour in Fig.~\ref{fig:order-two-daughter-wilson-loops}
        & \({\rm Re}\,\widetilde\chi\)
        & \({\rm Im}\,\widetilde\chi\)
        & identified sector
        \\
        \hline
        blue (lower core)
        & \(1.03\)
        & \(-3.33 \times 10^{-16}\)
        & \(D_2\) or \(D_3\)
        \\
        red (upper core)
        & \(1.03\)
        & \(-3.33 \times 10^{-16}\)
        & \(D_2\) or \(D_3\)
        \\
        black (both cores)
        & \(2.32 \times 10^{-3}\)
        & \(-1.22 \times 10^{-15}\)
        & \(D_6\)
        \\
    \end{tabular}
    \caption{
        Values of the lifted-holonomy trace \(\widetilde\chi\) evaluated along the three contours in Fig.~\ref{fig:order-two-daughter-wilson-loops}.
    }
    \label{tab:order-two-daughter-wilson-results}
\end{table}

For both local contours, $\widetilde{\chi}$ is close to $1$,
which, from Table~1, places each separated vortex in either
the $D_2$ or $D_3$ class.
By contrast, the contour surrounding both cores gives \(\widetilde\chi\simeq0\), corresponding to the \(D_6\) class of the original order-two \(N=1\) vortex.

The local trace invariants alone cannot distinguish \(D_2\) from \(D_3\), since these sectors correspond to the two orientations of a fundamental order-three vortex and have identical trace invariants.
The contour surrounding both vortices, however, shows that their total charge belongs to \(D_6\).
From Table~\ref{tab:binary-tetrahedral-fusion}, the \(D_6\) channel appears in \(\mathcal D_2\mathcal D_3\), but not in \(\mathcal D_2\mathcal D_2\) or \(\mathcal D_3\mathcal D_3\).
The two daughter vortices are therefore identified as one vortex in \(D_2\) and one in \(D_3\).
The observed single-core profiles and the energy approaching
$2E_1^{(3)}$ further support identifying these vortices with
the fundamental representatives $M=1$ and $M=-1$.
The total $\widetilde{A}_4$ topological charge is preserved.

\begin{figure}[htbp]
  \centering
  \includegraphics[width=1\textwidth]{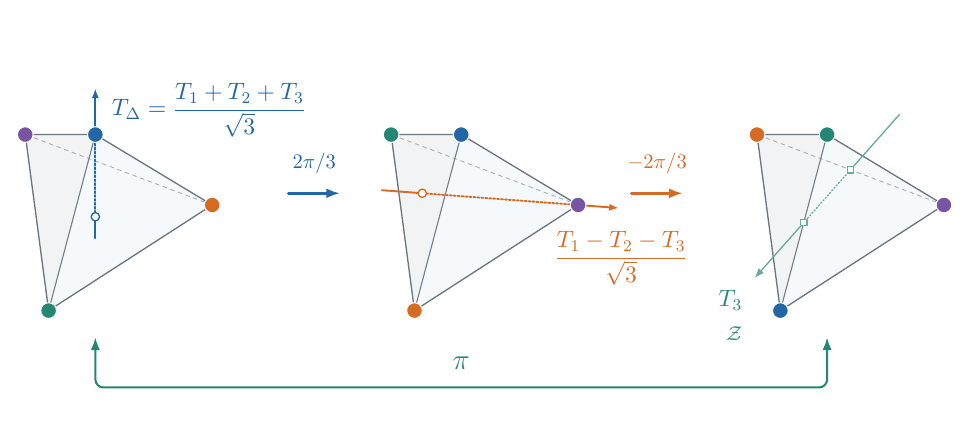}
  \caption{Composition of two order-three rotations into an order-two
rotation in the internal $(\mathcal X,\mathcal Y,\mathcal Z)$ space.
This provides an intuitive picture of how an order-two vortex
can decay into two order-three vortices.
The two right-pointing arrows show a $2\pi/3$ rotation about
the axis associated with $T_\Delta$, followed by a $-2\pi/3$
rotation about the axis associated with
$(T_1-T_2-T_3)/\sqrt{3}$.
Their composition equals the $\pi$ rotation about the
$\mathcal Z$-axis associated with $T_3$ shown by the lower double-headed arrow.
This is the $SO(3)$ projection of Eq.~\eqref{eq:two-three}.
The colours track the same vertices in all three panels.
The displayed angles $\alpha$ are defined by $R(\alpha)=e^{i\alpha T}$,
where $T$ is the generator indicated beside the axis; with the convention
in Eq.~\eqref{eq:SO(3)-generator}, the corresponding right-handed
rotation angle is $-\alpha$.}
  \label{fig:order-two-rotations}
\end{figure}

Finally, this decay has a simple geometric interpretation in terms
of tetrahedral rotations.
Fig.~\ref{fig:order-two-rotations} illustrates the representative
charge decomposition in Eq.~\eqref{eq:two-three}:
two order-three rotations about different internal axes combine
into an order-two rotation.

\section{Formation of an order-three vortex molecule}
\label{sec:order-three-molecule}

\subsection{Preliminary}
We now turn to the nonlinear fate of the unstable order-three \(M=2\) vortex.
As discussed in Sec.~\ref{subsec:review-linear-stability}, we consider the parameter point
\begin{align}
    (\kappa, \eta) = (\kappa_\star, \eta_\star)
    \simeq (0.56899, \,0.12068),
\end{align}
which lies in the region \(\Delta_{\rm bin}<0\) and \(\lambda_{\min}^{(2)}<0\).
At this point, our numerical calculation finds precisely two negative modes in the gauge-fixed Hessian evaluated at the axisymmetric \(M=2\) vortex.
These two modes are related by a spatial rotation and therefore represent the same splitting instability with different orientations in the transverse plane.

We investigate the nonlinear development of this instability by two independent gradient-flow calculations.
In the first calculation, we perturb the axisymmetric \(M=2\) vortex along one of its negative modes and follow the resulting flow.
In the second calculation, we construct an initial configuration consisting of two fundamental \(M=1\) vortices at finite separation and evolve it by the same gradient flow.
We find that the two flows approach the same finite-separation configuration, up to a gauge transformation.
This configuration therefore provides a vortex molecule composed of two fundamental order-three vortices.
The charge composition of this molecule has a similar geometric
interpretation in terms of tetrahedral rotations, as in
Fig.~\ref{fig:order-two-rotations}.
Here, the two constituent $M=1$ vortices correspond to $2\pi/3$
rotations about the same internal axis associated with $T_\Delta$.
Their composition is a $4\pi/3$ rotation with lifted charge
$\widetilde r_3^{\,2}$, matching that of the $M=2$ vortex.
We further analyze the stability of the vortex molecule and find that it is linearly stable.

For the nonlinear calculations in this section, we use the square domain
\begin{align}
    -10 \leq x, y \leq 10
\end{align}
with uniform lattice spacing
\begin{align}
    \Delta x = \Delta y = 0.04,
\end{align}
corresponding to \(501\times501\) lattice points, and integrate the gradient flow with
\begin{align}
    \Delta \tau = 2.0 \times 10^{-4}.
\end{align}
Both the order-three background and its unstable modes belong to the five-component sector
\begin{align}
    W_i = A_i T_\Delta, \qquad
    \phi = \phi_1 f_1 + \phi_6 f_6 + \phi_7 f_7,
\end{align}
so that only the five real fields \((A_x,A_y,\phi_1,\phi_6,\phi_7)\) need to be evolved.
As we show in the following subsection, this sector is closed under the full gradient flow and can be interpreted as an embedded Abelian gauge--Higgs system.

\subsection{Embedded Abelian sector}
\label{subsec:embedded-abelian-sector}

As noted in Sec.~\ref{subsec:review-linear-stability}, both the axisymmetric order-three vortex and its negative modes are contained in the five-component field sector
\begin{align}
    W_i = A_i T_\Delta,
    \qquad 
    \phi = u f_1 + \phi_6 f_6 + \phi_7 f_7,
    \label{eq:embedded-sector-fields}
\end{align}
where we have introduced \(u:=\phi_1\) for convenience in the Abelian description below.
This sector is closed under the full gradient flow:
if the initial configuration has the form \eqref{eq:embedded-sector-fields}, no other field components are generated during the flow.
To see this, consider the constant gauge transformation
\begin{align}
    r_3 = \exp\left(\frac{2\pi\ii}{3}T_\Delta\right).
\end{align}
Under its adjoint action, the fixed subspace of the Lie algebra \(\mathfrak{so}(3)\) is spanned by \(T_\Delta\).
For the Higgs field, Eq.~\eqref{eq:order-three-adapted-action} shows that the fixed subspace of \(r_3\) is \(\operatorname{span}\{f_1,f_6,f_7\}\).
Because the full energy is gauge invariant, its gradient is equivariant (and also invariant here) under \(r_3\), and hence the gradient evaluated within this sector contains only these five components.
Therefore, a configuration initially in this sector remains in it throughout the gradient flow.

It is also useful to introduce the complex scalar
\begin{align}
    \zeta := \phi_6 + \ii\phi_7.
\end{align}
Substituting Eq.~\eqref{eq:embedded-sector-fields} into the energy functional gives
\begin{align}
    E = \int d^2x \Bigg[
        \frac{1}{4} F_{ij}F_{ij} + \frac{1}{2}(\partial_i u)(\partial_i u)
        + \frac{1}{2} |\nabla_i \zeta|^2 
        + \widehat{U}(u, \zeta)
    \Bigg],
    \label{eq:embedded-energy}
\end{align}
where
\begin{align}
    F_{ij} &= \partial_iA_j-\partial_jA_i,\\
    \nabla_i\zeta &= (\partial_i-3\ii A_i)\zeta,\\
    \widehat{U}(u, \zeta) &= \frac{\kappa^2}{40\eta^2} \left(4u^2 - 5|\zeta|^2\right)^2 + \frac{\kappa^2}{2} \left(u^2 + |\zeta|^2 - 1\right)^2.
\end{align}
The form of Eq.~\eqref{eq:embedded-energy} shows that \(A_i\) can be regarded as an Abelian gauge field.
With respect to this gauge field, \(\zeta\) behaves as a complex scalar of charge three, while \(u\) behaves as a neutral real scalar.
The five-component sector can therefore be regarded as an embedded multicomponent Abelian gauge--Higgs model.

The finite-separation vortex molecule found below is similar to the type-1.5 behavior known in multicomponent Abelian-Higgs models.
In such systems, attractive and repulsive intervortex forces can compete, allowing vortices to form bound states at finite separation \cite{PhysRevB.72.180502,PhysRevB.83.174509}.
\footnote{
    In these works, both scalar fields are complex, whereas the neutral field \(u\) in the present embedded model is real.
}
Its stability with respect to fluctuations outside this sector is a separate question, which we address using the full \(SO(3)\) Hessian in Sec.~\ref{subsec:order-three-molecule-stability}.

For the second gradient-flow calculation described above, we need an initial configuration of two separated fundamental vortices.
We construct this configuration following the Abrikosov ansatz for well-separated Abelian-Higgs vortices \cite{Vilenkin:2000jqa}.
Let \(a(r)\), \(b(r)\), and \(J(r)\) denote the radial profiles of the \(M=1\) vortex, and define
\begin{align}
    r_\pm &= \sqrt{(x\mp d/2)^2 + y^2}, &
    \theta_\pm &= \arg\!\left(x\mp d/2 + \ii y\right),
\end{align}
where \(d\) is the prescribed vortex separation in the ansatz.
We take
\begin{align}
    A &= -\frac{1}{3} \left[J(r_+) d\theta_+ + J(r_-)d\theta_-\right],
    \label{eq:pair-ansatz-gauge} \\
    u &= \frac{\sqrt{5}}{3}\,b(r_+)b(r_-), \\
    \zeta &= \frac{2}{3}\,a(r_+)a(r_-)e^{-\ii(\theta_+ + \theta_-)}.
    \label{eq:pair-ansatz-zeta}
\end{align}
This ansatz places the two fundamental vortices at \((d/2,0)\) and \((-d/2,0)\).

To verify the local form near the vortex at \((d/2,0)\), set
\begin{align}
    X = x - \frac{d}{2},\qquad Y = y,
\end{align}
and introduce polar coordinates
\begin{align}
    X = \rho\cos\vartheta, \qquad Y = \rho\sin\vartheta.
\end{align}
Keeping \((\rho, \vartheta)\) fixed and taking \(d\to\infty\), we have
\begin{align}
    r_+ &\to \rho, &
    \theta_+ &\to \vartheta, &
    r_- &\to \infty, &
    \theta_- &\to 0.
\end{align}
Using the asymptotic values of the radial profiles \eqref{eq:order-three-boundary}, the configuration therefore approaches
\begin{align}
    A &\longrightarrow -\frac{1}{3}J(\rho)d\vartheta, \\
    u &\longrightarrow \frac{\sqrt{5}}{3}b(\rho), \\
    \zeta &\longrightarrow \frac{2}{3}a(\rho)e^{-\ii\vartheta}.
\end{align}
Since \(\zeta = \phi_6 + \ii\phi_7\), the last expression is equivalent to
\begin{align}
    \phi_6 &= \frac{2}{3}a(\rho)\cos\vartheta, &
    \phi_7 &= -\frac{2}{3}a(\rho)\sin\vartheta.
\end{align}
Together with the limiting forms of \(A\) and \(u\), this reproduces the isolated \(M=1\) vortex.
The same argument applies to the vortex centered at \((-d/2,0)\).

For finite \(d\), the configuration \eqref{eq:pair-ansatz-gauge}--\eqref{eq:pair-ansatz-zeta} does not in general satisfy the static field equations and is used only as an initial condition for the gradient flow.
In the limit \(d\to\infty\), however, the two vortices decouple and the static field-equation residual vanishes.
As shown in the initial configuration in Fig.~\ref{fig:order-three-molecule-formation}(c), the ansatz indeed contains two distinct vortex cores separated by a finite distance.

\subsection{Formation of the vortex molecule}

We now follow the gradient flow from two different initial configurations.
In the first calculation, we start from the axisymmetric \(M=2\) vortex with a small perturbation along one of its negative modes.
In the second, we start from the two-vortex configuration constructed in Sec.~\ref{subsec:embedded-abelian-sector} using the Abrikosov ansatz with \(d=5\).
In both cases, the flow settles into a two-core configuration with a finite separation.

\begin{figure}[p]
    \centering
    \includegraphics[width=0.95\textwidth]{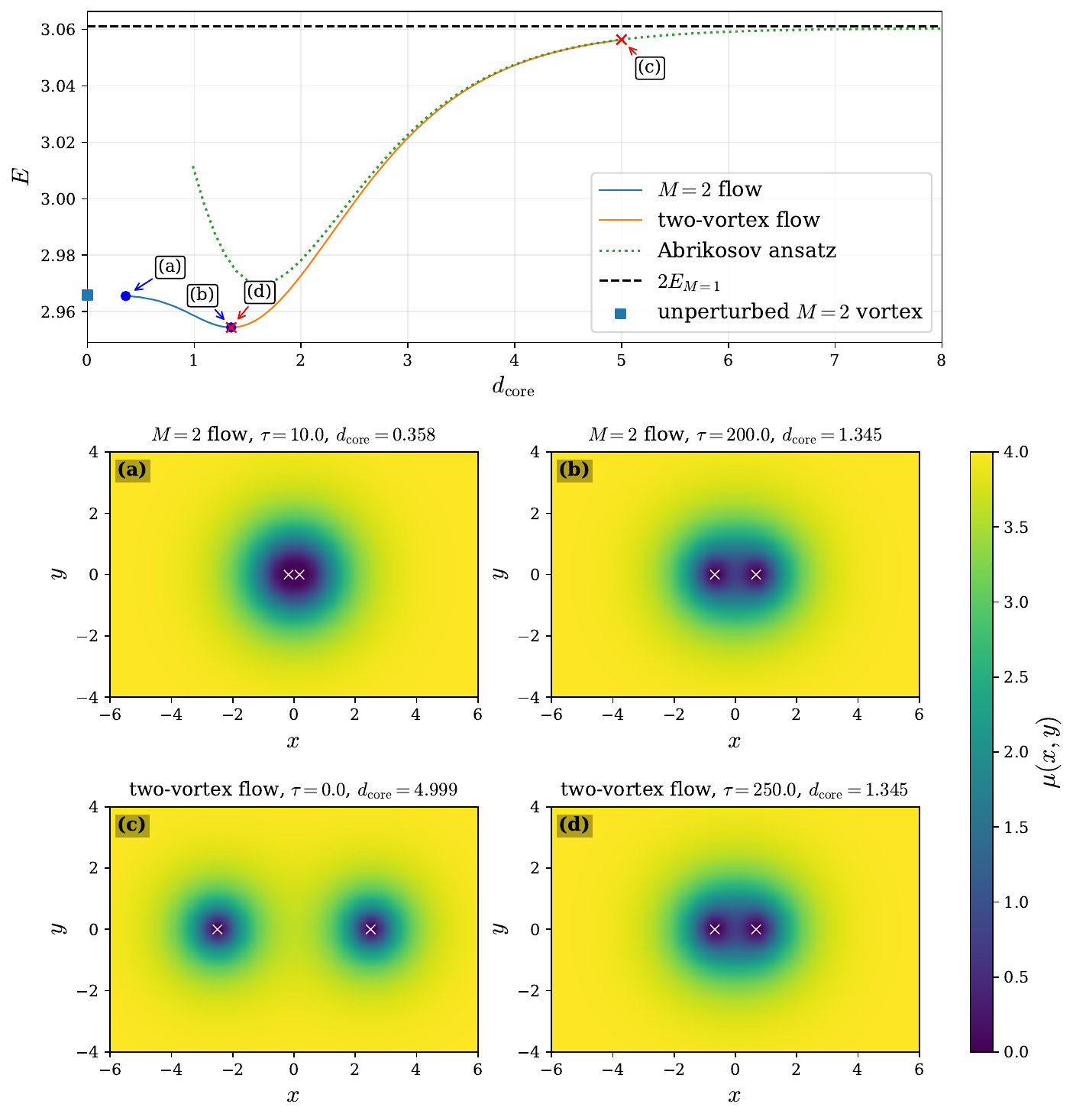}
    \caption{
        Formation of the order-three vortex molecule under the gradient flow.
        The upper panel shows the total energy as a function of \(d_{\rm core}\).
        The two flow trajectories start from the perturbed \(M=2\) vortex and from the separated two-vortex configuration.
        The horizontal dashed line gives twice the energy of an isolated \(M=1\) vortex, while the dotted curve shows the energy of the Abrikosov configurations.
        The labels (a)--(d) indicate the four configurations shown below:
        (a) the first resolved two-core configuration in the \(M=2\) flow,
        (b) the endpoint of the \(M=2\) flow,
        (c) the initial two-vortex configuration constructed with \(d=5\), and
        (d) the endpoint of the flow starting from the two separated \(M=1\) vortices.
    }
    \label{fig:order-three-molecule-formation}
\end{figure}

Fig.~\ref{fig:order-three-molecule-formation} summarizes the two flows.
The upper panel shows the total energy as a function of the gauge-invariant core separation \(d_{\rm core}\).
The two curves correspond to the flows starting from the perturbed \(M=2\) vortex and from the separated two-vortex configuration.
For comparison, we also show twice the energy of an isolated \(M=1\) vortex and the energy of the Abrikosov configurations constructed from the numerical \(M=1\) solution.
For the Abrikosov configurations, the prescribed separation \(d\) in the ansatz need not coincide exactly with the numerically measured core separation \(d_{\rm core}\); in Fig.~\ref{fig:order-three-molecule-formation}, each configuration is plotted at the latter, determined from the minima of \(\mu\).

Starting from the perturbed \(M=2\) vortex, the single core splits into two, as seen in the first resolved configuration in Fig.~\ref{fig:order-three-molecule-formation}(a).
The two cores then move apart and approach the finite-separation configuration shown in panel (b).
Starting from the separated configuration in panel (c), the two vortices instead move toward each other and approach the configuration shown in panel (d).
Thus, the two flows approach the molecular configuration from opposite sides in the energy--separation diagram.

The endpoints of the two flows, starting from the \(M=2\) vortex in panel (b) and from two separated \(M=1\) vortices in panel (d), have nearly identical core separations and energies:
\begin{align}
    d_{\rm core}^{(b)} &\simeq d_{\rm core}^{(d)} \simeq 1.345,\\
    E^{(b)} &\simeq E^{(d)} \simeq 2.954.
\end{align}
The relative differences in the core separation and energy are approximately \(2.2 \times 10^{-8}\) and \(3.4 \times 10^{-9}\), respectively.
This agreement provides further evidence that both flows approach the same vortex-molecule configuration.

To compare the configurations at the level of the fields, we next account for their relative \(U(1)\) gauge freedom.
Let
\begin{align}
    \Phi^{(\mathrm{M2})} &= \bigl(A_i^{(\mathrm{M2})}, u^{(\mathrm{M2})}, \zeta^{(\mathrm{M2})}\bigr), \\
    \Phi^{(\mathrm{pair})} &= \bigl(A_i^{(\mathrm{pair})}, u^{(\mathrm{pair})}, \zeta^{(\mathrm{pair})}\bigr)
\end{align}
denote the two final configurations.
For a gauge parameter \(\alpha\), we define
\begin{align}
    \Phi_{\alpha}^{(\mathrm{pair})} := \left(A_i^{(\mathrm{pair})} + \partial_i\alpha,\,u^{(\mathrm{pair})},\,e^{3\ii\alpha}\zeta^{(\mathrm{pair})}\right).
\end{align}
We determine their relative gauge by choosing \(\alpha\) to minimize
\begin{align}
    \delta[\alpha] = \int_{\Omega} d^2x\,
    \left|\Phi_{\alpha}^{(\mathrm{pair})} - \Phi^{(\mathrm{M2})}\right|^2,
\end{align}
where
\begin{align}
    |\Phi|^2 := \sum_i |A_i|^2 + |u|^2 + |\zeta|^2 .
\end{align}
We denote the minimizing gauge parameter by \(\alpha_\ast\).
To quantify the remaining difference between the two configurations, we define the relative field difference
\begin{align}
    \delta_{\rm rel} = \left[
    \frac{
        \displaystyle \int_\Omega d^2x\, \left|\Phi_{\alpha_\ast}^{(\mathrm{pair})} - \Phi^{(\mathrm{M2})}\right|^2
    }{
        \displaystyle \frac{1}{2} \int_\Omega d^2x\, \left(\left|\Phi_{\alpha_\ast}^{(\mathrm{pair})}\right|^2 + \left|\Phi^{(\mathrm{M2})}\right|^2\right)
    }
    \right]^{1/2}.
\end{align}
For the numerical evaluation, we take
\begin{align}
    \Omega = [-6,6] \times [-4,4].
\end{align}
We then find
\begin{align}
    \delta_{\rm rel} \simeq 1.2\times10^{-7} . 
\end{align}
We therefore conclude that the two gradient flows converge, up to a gauge transformation, to the same vortex molecule.

\subsection{Stability of the vortex molecule}
\label{subsec:order-three-molecule-stability}

The gradient-flow calculations above show that the unstable axisymmetric \(M=2\) vortex and a pair of separated \(M=1\) vortices both approach the same finite-separation configuration.
Since the nonlinear flow was performed within the five-component sector discussed in Sec.~\ref{subsec:embedded-abelian-sector}, it remains to determine whether the resulting vortex molecule is also stable against fluctuations in the full \(SO(3)\) theory.
For this purpose, we evaluate the full gauge-fixed Hessian \(\widetilde L\) introduced in Sec.~\ref{subsec:review-linear-stability} at the molecular configuration and numerically solve the corresponding eigenvalue problem on the two-dimensional plane.

The ten lowest eigenvalues of the gauge-fixed Hessian \(\widetilde L\) are summarized in Table~\ref{tab:molecule-lowest-eigenvalues}.
The three lowest eigenvalues are numerically very close to zero, whereas the remaining computed eigenvalues are positive and have substantially larger magnitudes than the first three.

\begin{table}[t]
    \centering
    \begin{tabular}{c c}
        \hline\hline
        eigenvalue & numerical value \\
        \hline
        \(\lambda_1\)    & \(-2.4 \times 10^{-3}\) \\
        \(\lambda_2\)    & \(-2.2 \times 10^{-3}\) \\
        \(\lambda_3\)    & \(-1.2 \times 10^{-3}\) \\
        \(\lambda_4\)    & \(0.20\) \\
        \(\lambda_5\)    & \(1.10\) \\
        \(\lambda_6\)    & \(1.25\) \\
        \(\lambda_7\)    & \(1.25\) \\
        \(\lambda_8\)    & \(1.39\) \\
        \(\lambda_9\)    & \(1.40\) \\
        \(\lambda_{10}\) & \(1.45\) \\
        \hline\hline
    \end{tabular}
    \caption{
        Ten lowest eigenvalues of the gauge-fixed Hessian \(\widetilde L\) around the vortex molecule.
        The first three are numerically close to zero, while the remaining computed eigenvalues are positive.
    }
    \label{tab:molecule-lowest-eigenvalues}
\end{table}

We now show that the three near-zero modes are the geometric zero modes associated with translations and spatial rotation of the molecule.
The two translational zero modes can be constructed directly from the molecular configuration in terms of gauge-covariant quantities \cite{Abe:2026grk} as
\begin{align}
    Z_x &= \left(G_{xi},\, \nabla_x\phi\right), \\
    Z_y &= \left(G_{yi},\, \nabla_y\phi\right).
\end{align}
Using the equations of motion, one can verify that these modes satisfy the background-gauge condition \eqref{eq:review-background-gauge}.
The rotational zero mode about the origin is similarly given by\footnote{
This gauge-covariant expression is obtained by adding the gauge variation
\((-\nabla_i\alpha,\alpha^aX_a\phi)\), with \(\alpha=R^jW_j\),
to the ordinary rotational variation
\( \left(
R^j\partial_jW_i+(\partial_iR^j)W_j,\;
R^j\partial_j\phi
\right)
\). Here, \(R^i \partial_i = \partial_\theta \).
}
\begin{align}
    Z_{\rm rot} = \left(
        R^j G_{ji},\,
        R^j \nabla_j \phi
    \right),
    \qquad
    (R^x, R^y) = (-y, x).
\end{align}
Unlike the translational modes, \(Z_{\rm rot}\) does not in general satisfy the background-gauge condition.
For the molecular configuration, we therefore determine a gauge zero mode \(Z_{\rm gauge}\) numerically so that
\begin{align}
    \widehat Z_{\rm rot} = Z_{\rm rot} + Z_{\rm gauge}
\end{align}
satisfies the background-gauge condition.

We then compare the space spanned by these three geometric zero modes,
\begin{align}
    \mathcal Z_{\rm geom} = \operatorname{span} \left\{Z_x,\, Z_y,\, \widehat Z_{\rm rot}\right\},
\end{align}
with the space spanned by the three near-zero eigenmodes of the Hessian,
\begin{align}
    \mathcal Z_{\rm Hess} = \operatorname{span} \left\{\Psi_1, \Psi_2, \Psi_3\right\},
\end{align}
where \(\Psi_a\) is the eigenmode corresponding to \(\lambda_a\) in  Table~\ref{tab:molecule-lowest-eigenvalues}.

To compare \(\mathcal Z_{\rm geom}\) and \(\mathcal Z_{\rm Hess}\), we first construct an orthonormal basis from the geometric zero modes.
Writing
\begin{align}
    (Z_1, Z_2, Z_3) = \left(Z_x,\, Z_y,\, \widehat Z_{\rm rot}\right),
\end{align}
we define
\begin{align}
    e_a = \sum_{b=1}^3 \left(\mathfrak{G}^{-1/2}\right)_{ab} Z_b, \qquad
    \mathfrak{G}_{ab} = \langle Z_a, Z_b\rangle.
\end{align}
By construction,
\begin{align}
    \langle e_a, e_b\rangle = \delta_{ab}.
\end{align}

Similarly, we denote by \(\{w_a\}_{a=1}^3\) an orthonormal basis of \(\mathcal Z_{\rm Hess}\).
We then define the overlap matrix
\begin{align}
    C_{ab} = \langle e_a, w_b\rangle .
\end{align}
Let \(\sigma_1, \sigma_2, \sigma_3\) be the singular values of \(C\).
They satisfy
\begin{align}
    0 \leq \sigma_a \leq 1,
\end{align}
and the two subspaces coincide if and only if
\begin{align}
    \sigma_1 = \sigma_2 = \sigma_3 = 1.
\end{align}

The resulting singular values are shown in Table~\ref{tab:molecule-zero-mode-subspace}.
All three values are close to 1, showing that \(\mathcal Z_{\rm geom}\) and \(\mathcal Z_{\rm Hess}\) agree within numerical accuracy.
\begin{table}[t]
    \centering
    \begin{tabular}{c c}
        \hline\hline
         & singular value \\
        \hline
        \(\sigma_1\) & \(1 - 5.8 \times 10^{-7}\) \\
        \(\sigma_2\) & \(1 - 6.1 \times 10^{-7}\) \\
        \(\sigma_3\) & \(1 - 2.3 \times 10^{-6}\) \\
        \hline\hline
    \end{tabular}
    \caption{
        Singular values of the overlap matrix between \(\mathcal Z_{\rm geom}\) and \(\mathcal Z_{\rm Hess}\).
        All three values are close to unity.
    }
    \label{tab:molecule-zero-mode-subspace}
\end{table}
We therefore identify the three near-zero eigenvalues in Table~\ref{tab:molecule-lowest-eigenvalues} with the two translational zero modes and the rotational zero mode of the vortex molecule.
Apart from these three geometric zero modes, no negative eigenvalues were found in the computed spectrum.
This provides strong numerical evidence for the linear energetic stability of the vortex molecule in the full \(SO(3)\) theory.

\section{Summary and Discussion}
\label{sec:summary-discussion}
In this work, we have investigated 
fission and molecules of non-Abelian $A_4$ vortices 
by studying 
the instabilities found in the linear analysis of Ref.~\cite{Abe:2026grk} using the gradient flow \eqref{eq:gradient-flow-gauge} and \eqref{eq:gradient-flow-Higgs}.
We have focused on two representative instabilities and found qualitatively different behaviors.

For the order-two \(N=1\) vortex at \(\kappa=\eta=1\), the perturbation along the negative mode has grown under the gradient flow, and the vortex has split into two fundamental order-three vortices.
The Wilson-loop analysis places the daughter vortices in
$D_2$ and $D_3$, while their core profiles and energy are
consistent with the fundamental representatives $M=1$ and $M=-1$.
After the two cores become resolved, their separation has increased throughout the range investigated by the gradient flow, while the energy has approached that of two isolated fundamental order-three vortices.

For the order-three \(M=2\) vortex, we have considered the parameter point
$(\kappa, \eta) =(\kappa_\star, \eta_\star)
\simeq (0.56899, 0.12068)$,
which lies in the region
$\Delta_{\rm bin} < 0, \;\lambda_{\min}^{(2)} < 0$.
Here, \(\Delta_{\rm bin}\) is the energy difference between the axisymmetric \(M=2\) vortex and two infinitely separated \(M=1\) vortices, while \(\lambda_{\min}^{(2)}\) is the lowest non-translational Hessian eigenvalue of the axisymmetric \(M=2\) vortex.
Thus, the coincident \(M=2\) vortex is locally unstable, even though its energy is lower than that of two infinitely separated fundamental vortices.
This combination led us in Ref.~\cite{Abe:2026grk} to conjecture that the instability may terminate at a finite vortex separation.

Our analysis has provided three pieces of evidence for such a vortex molecule.
First, starting from the axisymmetric \(M=2\) vortex with a small perturbation along its negative mode, the single core splits into two and approaches a configuration with a finite core separation.
Second, starting from two separated \(M=1\) vortices, the two cores move toward each other and approach the same configuration.
After accounting for the relative gauge transformation, the two endpoints agree at the level of the fields to a relative difference of order \(10^{-7}\).
Third, the resulting molecular configuration is linearly stable in the full \(SO(3)\) theory:
the three numerically near-zero modes are identified with the two translations and the spatial rotation of the molecule, 
while all the remaining computed Hessian eigenvalues are positive.

As shown in Sec.~\ref{subsec:embedded-abelian-sector}, the nonlinear evolution of the order-three vortex is contained in an exactly invariant five-component sector of the full theory.
Within this sector, the system reduces to an embedded Abelian gauge--Higgs model containing an Abelian gauge field, a charged complex scalar, and a neutral real scalar.
The formation of the finite-separation molecule is reminiscent of type-1.5 behavior in multicomponent Abelian-Higgs systems, where competing attractive and repulsive intervortex interactions can produce bound states at a preferred separation \cite{PhysRevB.72.180502,PhysRevB.83.174509}.

We now turn to several implications and open questions.
We first comment on the interaction between fundamental order-three vortices at the BPS-like point \(\kappa=\eta=1\).
As discussed in Ref.~\cite{Abe:2026zrk}, there exist static multi-vortex solutions with the fundamental vortex cores located at arbitrary positions,
as the case of BPS vortices in the Abelian-Higgs model \cite{Taubes:1979tm,Weinberg:1979er}.
Thus, two fundamental order-three vortices with \(M=1\), whose relative internal orientation is such that their fusion gives an \(M=2\) order-three vortex, experience no static force.
The situation changes for a pair with windings \(M=1\) and \(M=-1\).
When their based charges are mutual inverses, so that their fusion gives the vacuum sector, the interaction is attractive, as in an Abelian-Higgs vortex--antivortex pair.
The daughter vortices produced by the order-two decay have the same windings, \(M=1\) and \(M=-1\), but their fusion gives the nontrivial order-two sector.
In this case, the increasing separation observed in the gradient flow indicates a repulsive interaction.
Thus, even for the same pair of winding numbers, the interaction between fundamental order-three vortices depends on their relative internal orientation.

We next comment on the parameter dependence of the nonlinear evolution studied here.
For the order-two vortex, our calculation has been restricted to \(\kappa=\eta=1\), where the two daughter vortices appear to separate from each other under the gradient flow.
The linear stability of the order-two vortex depends on the mass ratios, so the decay found here need not occur throughout parameter space.
Moreover, even when the vortex splits into two fundamental order-three vortices, their interaction may change as the parameters are varied.
It is therefore possible that, for other values of \((\kappa,\eta)\), the splitting terminates at a finite separation and produces a molecular bound state.

A similar question arises for the order-three \(M=2\) vortex.
The overlap region identified above by \(\Delta_{\rm bin}\) and \(\lambda_{\min}^{(2)}\) provides a useful guide to where molecular solutions may occur.
It remains to be determined, however, whether the actual existence region of the molecule coincides with this overlap region.
Moreover, the existence of a molecular solution and its stability against
fluctuations outside the embedded Abelian sector are separate questions.
Determining the parameter regions in which molecular solutions exist and in
which they are nonlinearly stable in the full \(SO(3)\) theory remains an
important task for future work.

Another interesting direction is to investigate knotted and
linked configurations of non-Abelian cosmic strings.
Topological protection against allowed strand crossings and
reconnections has been established for certain non-Abelian
vortex knots and links~\cite{Annala:2022bdd,Rajamaki:2023ymv},
and their dynamical persistence has been studied in spinor
Bose--Einstein condensates~\cite{Kobayashi:2024aip}.
The present $SO(3)\to A_4$ model provides a gauge-theory setting
with smooth, finite-tension vortices carrying charges in the
binary tetrahedral group.
The fission and molecular binding found here motivate
investigating how the energetic stability and interactions of
constituent vortices affect the evolution and persistence of
knotted and linked configurations in three dimensions.

Finally, the gradient flow used in this work probes the static energy landscape and should not be identified with physical Lorentzian time evolution.
The flow time \(\tau\) is an auxiliary parameter, and the trajectories obtained here do not determine the physical decay time or the detailed dynamics of the vortices.
In particular, the increasing separation observed in the order-two flow does not by itself determine the real-time motion of the daughter vortices, while the existence of a stable order-three molecule does not determine how such a configuration would form dynamically.
Real-time simulations are required to study the decay of unstable vortices, the role of radiation and internal excitations, and the formation of molecular bound states in vortex scattering.
In cosmology, these dynamics may also be relevant to the evolution of non-Abelian cosmic-string networks.
A further extension to three-dimensional network simulations would be needed to understand their effects on network evolution and scaling behavior.

\section*{Acknowledgments}
\noindent
We developed the numerical codes used in this work with the assistance of ChatGPT (GPT-6, OpenAI).
ChatGPT was also used for English language editing of the manuscript.
The authors take full responsibility for the content of this work.
This work is supported by JST Grant No. JPMJPF2221 (YA),
by JST SPRING, Japan Grant No. JPMJSP2123 (KM),
by JSPS KAKENHI Grant No. JP22K03601 (TH),
JP22H01221 and JP23K22492 (MN), 
JP25K17394 (RY), and by 
the WPI program ``Sustainability with Knotted
 Chiral Meta Matter (WPI-SKCM$^2$)'' at Hiroshima University (MN).

\bibliographystyle{yautphysm}

\bibliography{draft.bib}
\end{document}